\documentclass{article}
\usepackage[utf8]{inputenc}
\usepackage[a4paper, margin=0.8in]{geometry}
\usepackage[style=nature,doi=false,isbn=false,eprint=false,url=false,maxbibnames=1001,]{biblatex}
\DeclareFieldFormat{sentencecase}{\MakeSentenceCase{#1}}
\AtEveryBibitem{\clearfield{month}}
\AtEveryBibitem{\clearlist{language}}
\AtEveryBibitem{\clearlist{urldate}}

\DeclareUnicodeCharacter{2212}{-}
\DeclareUnicodeCharacter{0327}{}

\usepackage{tabularx}
\usepackage{array}
\usepackage{amsmath}
\usepackage{amssymb}
\usepackage{booktabs}
\usepackage{makecell}
\usepackage{multirow}
\usepackage{tabulary}
\usepackage{xcolor}
\usepackage[normalem]{ulem}
\usepackage{ulem}
\usepackage[symbol]{footmisc}
\usepackage{siunitx}
\usepackage[verbatim]{minted}
\setminted[julia]{
    fontsize=\small,
    breaklines=true,
    autogobble=true,
    frame=lines,
    framesep=2mm,
    rulecolor=\color{black!20}
}
\usepackage{longtable}

\usepackage{graphicx}
\usepackage{caption}
\usepackage{subcaption}
\usepackage{float}
\usepackage{hyperref}
\usepackage[protrusion=true]{microtype}
\usepackage[capitalise,noabbrev]{cleveref}
\usepackage{fancyhdr}
\usepackage{authblk}

\title{PopulationBalanceModels.jl: An Extensible Framework for Deploying Population Balance Models}

\author[1]{Pavan K. Inguva*}
\affil[1]{Department of Chemical Engineering, Massachusetts Institute of Technology, 77 Massachusetts Avenue, Cambridge, MA 02139, USA}

\date{}

\begin{document}

\maketitle

\footnotetext[1]{Corresponding author. Email: inguvakp@mit.edu}
\renewcommand{\thefootnote}{\arabic{footnote}}

\begin{abstract}

Population balance models (PBMs) are used to describe a diverse range of processes that have a distribution over one or more intrinsic properties. However, implementing PBMs for a new application often requires developing and verifying application-specific code for numerical solution and analysis. In this work, we present \texttt{PopulationBalanceModels.jl}, an open-source Julia package designed to make PBMs easy to implement. Users construct models from individual physical terms and select the numerical methods separately. Common solver and observation interfaces allow the same model to be used for simulation, Bayesian parameter estimation, and uncertainty quantification. The package currently supports various model types such as coupled and/or multi-dimensional PBMs, coupling of a PBM to a bulk mass/energy balance, and transport in an axially-dispersed plug flow reactor. While a broad range of physical terms and corresponding numerical methods are already implemented, the package is designed to support extensibility in both aspects, enabling users to incorporate new mechanisms and numerical methods into existing workflows. 

\end{abstract}

\section{Introduction}

Population balance models (PBMs) are a powerful framework for describing the spatiotemporal dynamics of a population with a distribution over one or more intrinsic variables \cite{Ramkrishna2014}. A diverse range of intrinsic variables have been considered in the literature such as size (length / area/ volume), mass, age, composition, cellular characteristics, and catalytic activity, which has enabled PBMs to be applied to a wide range of systems including precipitation/crystallization \cite{Shin2025,Inguva2025,Omar2017,Schwarzer2006}, granulation / flocculation \cite{Yu2016,Kumar2016,Ahmad2008}, particle size reduction (e.g., emulsification / milling) \cite{Rijpert2026,Inguva2024,Lebaz2022}, biological cell populations \cite{Brown2025,Canova2022,Alhuthali2022,Sidorenko2008}, and bubbly flows \cite{Sanyal2004,Bhole2008}, among many others. PBMs can and often are extended in various ways such as including multiple intrinsic coordinates (e.g., crystal length and width) \cite{Orosz2025,Li2013}, coupling of the PBM with bulk balances e.g., concentration / temperature \cite{Srisuma2026,Oner2021}, introducing an axial coordinate to account for advection and dispersion along tubular reactors \cite{Semrau2024,BenitezChapa2019}, and coupling with computational fluid dynamics (CFD) to account for fluid flow \cite{Mukherjee2026,Marchisio2026,Woo2008}. 

In many cases, formulating a PBM to describe the relevant physical processes in a system of interest often results in a complex integro-partial differential equation with nonlinear terms and complex boundary conditions. This makes numerical solution necessary for most applications and has correspondingly motivated an extensive literature on developing efficient and accurate methods for PBMs (see Table~\ref{tab:pbm_numerics} for examples). However, the software landscape for PBMs remains fragmented across tools with different application scopes, numerical capabilities, and modeling interfaces. Commercial tools such gPROMS and PARSIVAL \cite{Wulkow2001} provide strong support for employing PBMs, but require commercial licenses and offer limited scope for users to inspect, modify, or extend the underlying numerical methods. PBM modules are also available within CFD and multiphysics solvers such as Ansys Fluent and COMSOL, facilitating coupling with spatial transport. Additionally, the PBMs in these systems are limited in their functionality such as supporting only a single intrinsic variable \cite{COMSOLPopulationBalance,FluentTheoryGuide}. PBM solvers have also been incorporated into OpenFOAM \cite{Heylmun2019,daRosa2018}, but their close integration with specific flow applications complicates their adaptation to other PBM problems. Open-source PDE solver codes such as FiPy \cite{Guyer2009}, Clawpack \cite{Mandli2016}, or MOOSE \cite{Wiechert2023,Gaston2009} have been used to solve PBMs, but often require significant user effort on to implement and validate PBM-specific requirements such as source/sink terms and specific numerical methods. More recently, native PBM functionality is also available in open-source process/application specific packages such as PharmaPy \cite{Shahab2026,CasasOrozco2021} and CADET \cite{Zhang2025,Zhang2024}, with PBM implementations tailored to particular unit operations and population formulations. Extending these packages would require additional implementation. In many cases, users often develop custom code for their specific application, which increases adoption barriers for PBM-based workflows and often leads to redundant development efforts for similar numerical components. 

This work presents \texttt{PopulationBalanceModels.jl}, an open-source Julia package for developing and deploying PBMs. Its current functionality includes simulation, parameter estimation, uncertainty quantification, and basic visualization. The package aims to enable the rapid and accessible implementation of PBMs across a wide range of applications, while user control over modeling methodology and numerical methods. By providing reusable and verified numerical and analysis tools, the package also seeks to reduce the programming effort for individual applications and encourage shared development by users. The Julia programming language was selected as it combines high-level programming with efficient compiled numerical execution, allowing model definition and computationally intensive methods to be developed together. Multiple dispatch supports code reuse and extensibility by selecting specialized implementation through common interfaces according to argument types \cite{Bezanson2017}. Support for automatic differentiation allows derivatives of models to be evaluated efficiently for sensitivity analysis and parameter estimation \cite{Revels2016}. Julia also has a robust ecosystem of packages that provide performant tools for functions such as ODE solvers \cite{Rackauckas2017}, nonlinear equation solvers \cite{Pal2026}, and optimization \cite{KMogensen2018}. 

The rest of the manuscript is structured as follows: Section 2 outlines the package architecture, model declarations, numerical methods, and other core functionality. Code examples based on representative PBMs are also presented. Section 3 presents four case studies based on previously published works to demonstrate the functionality of the package for a wide range of PBM types. Concluding remarks and a brief roadmap for future developments are presented in Section 4. 

\section{Package Description and Capabilities}

\texttt{PopulationBalanceModels.jl} provides tools for defining and solving PBMs, estimating model parameters, and propagating uncertainty in model inputs. The package separates the formulation of the PBM (i.e., the mechanisms and kinetic expressions incorporated) from the numerical methods used for solution. Both direct discretization of the PBM, and moment-based methods are supported. Separate interfaces can be used couple the PBM with bulk balances or incorporate prescribed reactor transport, and translate model predictions into experimentally measured quantities such as size distributions and concentrations. 

\subsection{Design Principles and Architecture}

\begin{figure}[h!]
\centering
\includegraphics[width=0.75\textwidth]{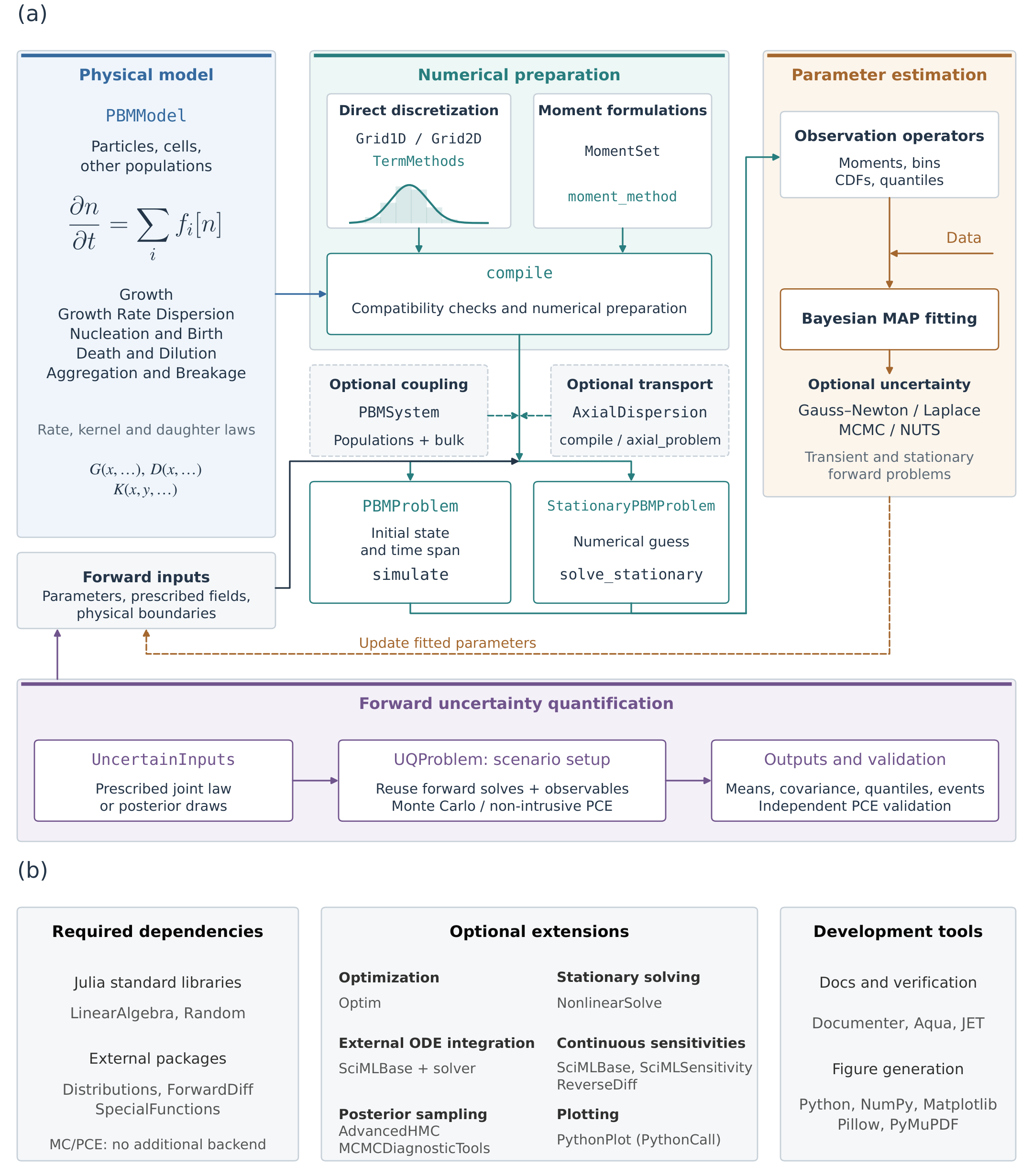}
\caption{Architecture and computational workflow of \texttt{PopulationBalanceModels.jl}. (a) Physical model terms are combined with a grid or moment representation and associated numerical methods to construct transient or stationary forward problems. This can optionally include coupled populations, bulk balances, and transport in an axially-dispersed plug flow reactor. Observation operators connect predictions to experimental data for parameter estimation via MAP. Parameter uncertainty can be assessed using Gauss--Newton or Laplace approximations or Markov-chain Monte Carlo (MCMC) sampling. Forward uncertainty quantification either by Monte Carlo simulation or non-intrusive polynomial chaos expansion (PCE) reuse the models and observables. (b) Required runtime dependencies, optional Julia extensions, and development tools for documentation, verification, and figure generation.}
\label{fig:architecture}
\end{figure}

Inspired by the equation-composition approach of FiPy \cite{Guyer2009}, a model in \texttt{PopulationBalanceModels.jl} is defined by specifying the terms that describe the evolution of a population distribution for a given application. Each term explicitly identifies the process it represents, such as growth, birth, death, aggregation, or breakage, and is supplied with the associated rate functions, kernels, and/or other required physical inputs. The complete model is constructed by combining the terms. This keeps model definition close to the structure of the governing equation, making the contribution of each process clear in the code. Such a structure can be beneficial for users with minimal coding experience looking to implement modeling workflows \cite{Inguva2021}.

Model definition is kept separate from the numerics. Users can solve the PBM through direct discretization of the PDE on a grid, or through its moment equations. The numerical setup specifies the methods applied to individual terms, the moment formulation and any required closure, and the time discretization. The separation allows the numerical schemes to be chosen according to the nature of the problem, accuracy requirements, and computational budget. This flexibility is particularly useful for computationally-demanding workflows such as parameter estimation or model predictive control. Existing methods can be reused across applications, while new kinetic laws and numerical methods can be incorporated through the package interfaces. This reduces the need for application-specific solver implementations and provides shared components for verification and maintenance. 

The model definition and selected numerical methods are brought together through \texttt{compile}, which checks compatibility, selects the routines for evaluating each term, and precomputes reusable coefficients, such as grid-dependent reconstruction weights. Initial conditions, parameter values, and the simulation time span are supplied through \texttt{PBMProblem}. In cases where the PBM involves coupled populations and/or scalar bulk balances, \texttt{PBMSystem} assembles the model into a single system of equations. Rate functions and interaction terms are evaluated from the same state at each integration stage, supporting the coupling between bulk variables and populations. For axially-dispersed plug flow reactors, an \texttt{AxialDispersion} domain adds prescribed advection and dispersion terms through \texttt{compile} for standalone directly discretized populations, or through \texttt{axial\_problem} for eligible coupled local problems. Transient problems are solved using \texttt{simulate}, which automatically invokes \texttt{prepare\_run} to prepare parameter-dependent numerical coefficients and working storage for each run or operating interval. Supported steady-state models use \texttt{StationaryPBMProblem} and \texttt{solve\_stationary}. 

Observation operators can be constructed to map simulated states to experimentally measured quantities, providing the model predictions used in evaluating the likelihood for Bayesian parameter estimation. The package supports maximum a posteriori (MAP) estimation, local parameter uncertainty analysis using Gauss--Newton or Laplace approximations, and posterior sampling using Markov chain Monte Carlo (MCMC). These methods reuse the forward model as parameter values vary. The same model also supports forward uncertainty propagation and quantification using Monte Carlo simulation or non-intrusive polynomial chaos expansion (PCE). The complete workflow is summarized in Fig.~\ref{fig:architecture}a. 

The simulation core uses Julia's \texttt{LinearAlgebra} and \texttt{Random} standard libraries, together with \texttt{Distributions}, \texttt{ForwardDiff}, and \texttt{SpecialFunctions}. Optional Julia packages provide the functionality for optimization for MAP estimation, external ODE integration, stationary solvers, continuous sensitivity analysis, MCMC posterior sampling, and plotting. Documentation, verification, and figure generation use separate development tools. The required dependencies, optional packages, and development tools are summarized in Fig.~\ref{fig:architecture}b. 

\subsection{Numerical Methods}

The numerical setup specifies the population representation (i.e., direct discretization or moments), and the methods applied to individual model terms. Per-term treatments are supplied through \texttt{TermMethods}, either by mechanism or through overrides for named contributions. In the following example, the MUSCL scheme for the growth term and fixed-pivot deposition for nucleation is selected. A reusable \texttt{ForwardSolver} specifies the time integrator and execution settings, in this case SSPRK33 with a maximum timestep of 0.01, and is passed to \texttt{simulate}.
\begin{minted}{julia}
methods = TermMethods(growth=MUSCL(), nucleation=FixedPivot())
operator = compile(model, grid; methods, bc)
problem = PBMProblem(operator, initial, (0.0, t_final); p)

solver = ForwardSolver(SSPRK33(); dt=0.01)
solution = simulate(problem; solver)
\end{minted}

For moment formulations, \texttt{MomentSet(0:K)} specifies the moments to be tracked, from the zeroth moment up to order $K$ for one internal coordinate, while \texttt{compile(...; moment\_method=...)} selects the moment method. The compiler uses implemented analytical moment expressions for supported term and rate-law declarations including \texttt{PolynomialRate} expressions. The selected moment method approximates any untracked moments required by these expressions and evaluates contributions without analytical formulas by quadrature. \texttt{ExactClosure} requires all contributions to have analytical expressions involving only tracked moments and is the default when this condition is met and the moment boundary conditions are compatible. Time integration remains numerical. 

For transient problems, \texttt{simulate} applies the selected time integrator to the assembled population and bulk equations. \texttt{ForwardSolver} centralizes integration settings and execution checks, reducing application-specific solver code and enabling reuse across different workflows. Complete space-time growth updates incorporate their own temporal discretization. The first-order \texttt{MeasureEuler} policy uses positive-measure splitting for eligible configurations, with admissibility checks and timestep rejection. For steady-state problems, \texttt{solve\_stationary} solves the system of algebraic equations using the \texttt{RobustMultiNewton} solver in \texttt{NonlinearSolve.jl} by default \cite{Pal2026}. The principal numerical methods are summarized in Table~\ref{tab:pbm_numerics}.

\begin{table}[ht!]
\centering
\small
\setlength{\tabcolsep}{2.5pt}
\renewcommand{\arraystretch}{1.1}

\caption{Numerical methods currently available in
\texttt{PopulationBalanceModels.jl}, grouped by numerical role. Growth entries distinguish spatial discretizations
from complete space--time updates. Supported combinations depend on
the representation, model terms, mesh and boundary conditions.
References identify underlying method families and relevant
methodological background.}
\label{tab:pbm_numerics}

\begin{tabularx}{\linewidth}{
    @{}
    >{\raggedright\arraybackslash}p{0.11\linewidth}
    >{\raggedright\arraybackslash}p{0.18\linewidth}
    >{\raggedright\arraybackslash}X
    >{\raggedright\arraybackslash}p{0.3\linewidth}
    >{\raggedright\arraybackslash}p{0.1\linewidth}
    @{}
}
\toprule
Numerical task
& Application
& Schemes and methods
& Notes
& References \\
\midrule

% ------------------------------------------------------------
% PDE discretization
% ------------------------------------------------------------

\multirow[t]{5}{=}{PDE discretization}

& \texttt{Growth}, \texttt{Growth2D}:
spatial fluxes; axial advection

& \texttt{Upwind}, \texttt{MUSCL}, \texttt{VanLeer},
\texttt{WENO3}, \texttt{WENO5}, \texttt{Centered},
\texttt{LinearUpwind}.

& Available 2D growth fluxes are \texttt{Upwind} and
\texttt{WENO5}. \texttt{Centered} and \texttt{LinearUpwind}
require uniform grids. Axial choices depend on the adapter.

& \cite{vanLeer1979,Jiang1996,Zhang2024} \\

\cmidrule(lr){2-5}

& \texttt{Growth}, \texttt{Growth2D}:
complete space--time updates

& \texttt{LaxWendroff}, \texttt{BeamWarming},
\texttt{ExactShift}, \texttt{Inguva2022},
\texttt{InguvaBraatz2023}, \texttt{HOCFiniteVolume},
\texttt{HOCADI}.

& Include temporal discretization.
\texttt{HOCFiniteVolume} is a conservative compact adaptation;
\texttt{HOCADI} is restricted to periodic constant-growth
2D transport. Other restrictions are method-specific.

& \cite{Zhang2025HOC,Inguva2023,Inguva2022,Leveque2002} \\

\cmidrule(lr){2-5}

& \texttt{SizeDiffusion}, Axial dispersion

& \texttt{FiniteVolumeDiffusion}.

& \texttt{FiniteVolumeDiffusion} applies to dispersion in the intrinsic coordinate, with diagonal coefficients in 2D.
Axial dispersion uses a separate centered flux assembled by
\texttt{AxialDispersion}.

& --- \\

\cmidrule(lr){2-5}

& \texttt{Birth} and \texttt{Nucleation} families;
\texttt{Death}, \texttt{Dilution}

& \texttt{MidpointRule}, \texttt{GaussLegendre},
\texttt{BinIntegral}, \texttt{FixedPivot},
\texttt{BoundaryInjection}, \texttt{GaussianInjection},
\texttt{PointwiseLoss}, \texttt{CellAverageLoss}.

& Quadrature, bin integrals or deposition for sources;
pointwise or cell-averaged removal hazards.
\texttt{GaussianInjection} numerically regularizes point sources.

& \cite{Leveque2002,Kumar1996} \\

\cmidrule(lr){2-5}

& \texttt{Aggregation}, \texttt{Breakage}

& \texttt{FixedPivot}.

& Fixed-pivot event counting and product deposition. Checked conservative projection for \texttt{ContinuousDaughters}.

& \cite{Kumar1996} \\

% ------------------------------------------------------------
% Moment formulations
% ------------------------------------------------------------

\midrule

\multirow[t]{3}{=}{Moment formulations}

& Supported closed moment systems

& \texttt{ExactClosure}.

& Analytical evaluation of all contributions using only tracked moments with compatible boundary conditions.

& --- \\

\cmidrule(lr){2-5}

& Quadrature-based formulations

& \texttt{QMOM}, \texttt{RankAdaptiveQMOM},
\texttt{EQMOM}, \texttt{DQMOM}.

& \texttt{EQMOM} supports Gamma kernels.
\texttt{DQMOM} evolves weights and weighted abscissas.

& \cite{Yuan2012,Marchisio2005,McGraw1997} \\

\cmidrule(lr){2-5}

& Assumed-distribution closure

& \texttt{LogNormalClosure}.

& Assumes a lognormal distribution using moments
of orders $0$, $1$ and $2$.

& \cite{Lee1983} \\

% ------------------------------------------------------------
% Time integration
% ------------------------------------------------------------

\midrule

\multirow[t]{2}{=}{Time integration}

& Assembled population and bulk residuals

& \texttt{ForwardEuler}, \texttt{SSPRK22},
\texttt{SSPRK33}, \texttt{SSPRK54}, \texttt{RK4}.

& Explicit residual integration.
Stability and positivity depend on the spatial
discretization and timestep restrictions.

& \cite{Spiteri2002,Shu1988} \\

\cmidrule(lr){2-5}

& Realizability-preserving evolution

& \texttt{MeasureEuler}.

& First-order positive-measure Lie splitting
for eligible configurations.

& \cite{Nguyen2016} \\

\bottomrule
\end{tabularx}
\end{table}

\subsection{Representing a Physical Model}

Consider a population described by volume, $v \in [0, v_{\max}]$, as the intrinsic coordinate, undergoing size-dependent growth, nucleation, and first-order removal in a spatially well-mixed system. The PBM can be written as
\begin{equation}
\begin{aligned}
\frac{\partial n}{\partial t} + \frac{\partial (G(v)n)}{\partial v} = J\delta(v-v_{b}) -k_{d}n, &\quad v\in(0,v_{\max}), \ t \in (0, t_{f}], \\
n(v,0) = 0, &\quad v\in[0,v_{\max}], \ t = 0, \\
G(0)n(0,t) = 0, &\quad v = 0, \  t\in(0,t_{f}],\\
\mathrm{Free  \ outflow}, &\quad v = v_{\max}, \ t\in(0,t_{f}],
\end{aligned}
\end{equation}
where $n$ is the number density, $J$ is the nucleation rate at the birth volume $v_{b}$ and is assumed constant, $k_{d}$ is the removal hazard rate constant, and $G$ is the growth rate and is given by $g_{0}(1+\alpha v)$ where $g_{0}$ and $\alpha$ are constants. All the model parameters are taken as positive and their illustrative values are specified in the code below. A zero incoming number flux boundary condition is applied at $v=0$ and a free outflow boundary condition is applied at $v=v_{\max}$. 

Parameter values are supplied separately from the model terms. Here, the kinetic parameters are stored in a Julia \texttt{NamedTuple} and the birth volume, upper volume limit and final simulation time are defined alongside them:
\begin{minted}{julia}
using PopulationBalanceModels

p = (g0=0.2, alpha=0.1, J=0.05, kd=0.02)
v_birth, v_max, t_final = 0.1, 10.0, 10.0
\end{minted}

The kinetic expressions are declared as Julia functions:

\begin{minted}{julia}
growth_rate(v, p, t, ctx) = p.g0 * (1 + p.alpha * v)
nucleation_rate(p, t, ctx) = p.J
removal_rate(v, p, t, ctx) = p.kd
\end{minted}

Kinetic functions receive the parameter container \texttt{p}, the current time \texttt{t}, and a context \texttt{ctx} as arguments. The required argument list depends on the term type with growth and removal functions additionally receiving the internal coordinate, even when the expressions do not depend on it. The same interfaces accommodate time-dependent conditions and inputs from coupled states. When the corresponding bulk variables / prescribed fields have been supplied, a growth rate can for example, access the bulk concentration $c$ through \texttt{ctx.bulk.c} and prescribed temperature through \texttt{ctx.fields.T}. Each kinetic function is supplied to its corresponding physical term constructor, and the term objects are combined into a \texttt{PBMModel}:
\begin{minted}{julia}
model = PBMModel(
    growth = Growth(rate=growth_rate),
    primary = Nucleation(rate=nucleation_rate, size=v_birth),
    loss = Death(rate=removal_rate),
)
\end{minted}

The \texttt{PBMModel} constructor accepts keyword arguments, whose names in this example (\texttt{growth},\texttt{primary},\texttt{loss}) serve as user-selected labels for individual mechanisms to be captured by the PBM. Their values are term objects whose types identifies the physical processes represented. The arguments supplied to each term constructor specify its functional form and other physical inputs. The available model terms and source constructors are summarized in Table~\ref{tab:pbm_terms}.

\begin{table}[h!]
\centering
\small
\caption{Mechanistic model terms commonly used in PBMs and their constructors in
\texttt{PopulationBalanceModels.jl}. Supported combinations depend on the
representation and numerical method.}
\label{tab:pbm_terms}
\begin{tabularx}{\linewidth}{
    p{0.22\linewidth}
    p{0.27\linewidth}
    X
}
\toprule
Mechanism & Constructors & User-supplied input \\
\midrule

Growth
& \texttt{Growth}, \newline \texttt{Growth2D}
& Growth rate function in each internal coordinate. \\

Growth rate dispersion
& \texttt{SizeDiffusion}
& Nonnegative dispersion coefficient in 1D or a pair of diagonal coefficients in 2D supplied as a
constant or function. \\

Birth sources
& \texttt{Birth} \newline \texttt{IntegratedBirth} \newline
\texttt{MomentSource}
& Source density, bin-integrated source, or creation rate together with raw moments per created population member. \\

Nucleation
& \texttt{Nucleation} \newline \texttt{Nucleation2D} \newline \texttt{DistributedNucleation} \newline \texttt{IntegratedNucleation} \newline \texttt{GaussianNucleation}
& Creation rate and birth coordinates or birth distribution. The distribution can be supplied through its density or bin integrals. Gaussian sources use a center and standard deviation. \\

Removal and dilution
& \texttt{Death} \newline \texttt{Dilution}
& Nonnegative removal hazard or concentration-dilution rate. \\

Aggregation
& \texttt{Aggregation}
& Symmetric, nonnegative binary aggregation kernel. \\

Breakage
& \texttt{Breakage}
& Parent breakage hazard and daughter coordinates with their expected counts, or a continuous daughter number density declared through \texttt{ContinuousDaughters}. \\

Custom contributions
& \texttt{CustomTerm}
& An in-place function that adds to the population derivative. \\

\bottomrule
\end{tabularx}
\end{table}

The internal-coordinate boundary conditions are specified using \texttt{SizeBC}, with the lower boundary first. For the positive growth rate considered here, \texttt{NoFlux()} implements zero incoming number flux at the lower boundary while \texttt{OpenBoundary()} permits outward transport at the upper boundary.
\begin{minted}{julia}
bc = SizeBC(NoFlux(), OpenBoundary())
\end{minted}

To construct the simulation problem, the model is paired with a numerical representation of the population. Direct PDE discretization uses \texttt{Grid1D} or \texttt{Grid2D} for one or two internal coordinates. The grids are specified by bin-face coordinates along each axis, while state entries store bin-integrated number concentrations. In this example, a uniform volume grid with 100 bins (101 bin edges) is defined and passed to \texttt{compile} together with the model and boundary conditions:
\begin{minted}{julia}
grid = Grid1D(
    range(0.0, v_max; length=101);
    coordinate=:volume,
)

operator = compile(model, grid; bc)
\end{minted}
The initial conditions in this case are represented by a vector of zeros. The full problem is defined using the \texttt{PBMProblem} constructor which takes in the compiled model, initial conditions, simulation time span, and model parameters:
\begin{minted}{julia}
initial = zeros(length(grid))

problem = PBMProblem(
    operator,
    initial,
    (0.0, t_final);
    p,
)
\end{minted}
For supported representations, \texttt{PBMProblem} also accepts \texttt{InitialDensity} or \texttt{InitialDistribution} which constructs the initial state from a number density or probability distribution respectively with a specified total number concentration. 

Steady-state problems use the same physical model, boundary conditions, and compiled PBM, with the time derivatives set to zero to obtain a system of algebraic equations. The \texttt{StationaryPBMProblem} constructor takes the compiled PBM, a numerical guess for the solution, and the model parameters. The option \texttt{autonomous=true} declares time-independent rate functions and operating conditions. The resulting problem is solved using \texttt{solve\_stationary} with the use of \texttt{NonlinearSolve.jl}.

\subsection{Coupled Populations}

Coupled PBMs describe processes containing distinct populations whose evolution is linked through interaction, transitions between populations or shared bulk quantities. As an illustrative example, consider a predator-prey model with age, $a\in[0,a_{\max}],$ as the intrinsic coordinate. The prey and predator age distributions are denoted by $n(a,t)$ and $m(a,t)$ respectively, with total concentrations,
\begin{equation}
    X(t) = \int_{0}^{a_{\max}}n(a,t) da, \quad Y(t) = \int_{0}^{a_{\max}} m(a,t) da. 
\end{equation}
A coupled PBM describing birth and death of the two populations with age-independent coefficients is
\begin{equation}
\begin{aligned}
\frac{\partial n}{\partial t} + \frac{\partial n}{\partial a} &= -kY(t)n, \quad n(0,t) = rX(t), \\
\frac{\partial m}{\partial t} + \frac{\partial m}{\partial a} &= -dm, \quad m(0,t) = \eta kX(t)Y(t),
\end{aligned}
\end{equation}
where $r$ is the prey birth-rate constant, $k$ is the predation coefficient, $d$ is the predator mortality rate, and $\eta$ is a factor that converts prey consumption into predator births. The rate functions access the current population totals through the shared context \texttt{ctx}:
\begin{minted}{julia}
total(ctx, name) = moment(ctx.system, name, 0)
aging(a, p, t, ctx) = 1.0

prey_birth(p, t, ctx) = p.r * total(ctx, :prey)
prey_death(a, p, t, ctx) = p.k * total(ctx, :predators)

predator_birth(p, t, ctx) =
    p.eta * p.k * total(ctx, :prey) * total(ctx, :predators)

predator_death(a, p, t, ctx) = p.d
\end{minted}
Each population is declared separately using \texttt{Growth} for aging with unit velocity and \texttt{Death} for removal:
\begin{minted}{julia}
prey_model = PBMModel(
    aging=Growth(rate=aging),
    predation=Death(rate=prey_death),
)

predator_model = PBMModel(
    aging=Growth(rate=aging),
    mortality=Death(rate=predator_death),
)  
\end{minted}

Births enter through the zero-age boundary using \texttt{InflowFlux}. The models are compiled with their associated boundary conditions and age grid:
\begin{minted}{julia}
age_grid = Grid1D(
    range(0.0, a_max; length=101);
    coordinate=:age,
)

prey_bc = SizeBC(InflowFlux(prey_birth), OpenBoundary())
predator_bc = SizeBC(InflowFlux(predator_birth), OpenBoundary())

prey_operator = compile(prey_model, age_grid; bc=prey_bc)
predator_operator = compile(predator_model, age_grid; bc=predator_bc)
\end{minted}

\texttt{PBMSystem} is used to assemble the compiled population balances into a compiled system of equations while \texttt{PBMProblem} is used to specify the intial conditions, model parameters, and simulation timespan:
\begin{minted}{julia}
system = PBMSystem((
    prey=prey_operator,
    predators=predator_operator,
))

problem = PBMProblem(
    system,
    (prey=prey_initial, predators=predator_initial),
    (0.0, t_final);
    p,
)
\end{minted}
The coupled problem is then solved using \texttt{simulate}. In this example, rate and boundary callbacks evaluate the populations' zeroth moments from the shared states supplied at each integration stage to determine the interaction rates and boundary fluxes. Each population can use its own grid, moment formulation, and numerical settings if specified. Shared scalar bulk balances can also be included and integrated as part of the same system. 

\subsection{Axially-Dispersed Plug Flow Reactors}

Populations undergoing change in an axially-dispersed plug flow reactor can be modeled with advection and dispersion along a single physical reactor coordinate $z$. \texttt{AxialGrid} defines the axial mesh separately from the internal-coordinate representation and \texttt{AxialDispersion} specifies the prescribed constant velocity, axial dispersion coefficient, and population distribution at the reactor inlet. Local kinetic functions access the axial position through \texttt{ctx.fields.z} and user-imposed inputs such as a temperature profile through \texttt{ctx.fields}. 

For direct PDE discretization, the existing model, internal-coordinate grid and numerical settings are combined with the reactor domain in a single \texttt{compile} call. The numerical scheme for the advection term is selected independently of the internal-coordinate growth scheme through \texttt{axial\_scheme}:
\begin{minted}{julia}
zgrid = AxialGrid(range(0.0, 1.0; length=51))
reactor = AxialDispersion(
    zgrid;
    velocity=0.1,
    dispersion=0.001,
    inlet=nothing,
)

reactor_operator = compile(
    model, grid;
    domain=reactor,
    axial_scheme=MUSCL(),
    methods,
    bc,
)
\end{minted}
The default \texttt{inlet=nothing} specifies zero population concentration in the reactor feed. A continuous feed number density profile (e.g., continuous seeding) can be supplied using \texttt{DensityInlet(density)}, which integrates \texttt{density(v,p,t)} over the represented internal-coordinate bins. Reactor boundary conditions are specified through \texttt{AxialDispersion}, separately from the internal-coordinate boundary conditions, and default to a Danckwerts inlet condition and advective outflow with zero dispersive flux at the outlet.

\subsection{Parameter Estimation}
\label{sec:param_estimation}

A Bayesian estimation approach is used for parameter estimation tasks. The measured data that can be used for parameter estimation include measurements of the population distribution, its moments, or other bulk quantities such as concentrations depending on the process being studied. The forward problem, i.e., the numerical solution of the PBM as described above provides predictions for candidate parameter values. Observation operators are then used to map the simulated states to the corresponding measured quantities, accounting for experimental binning / weighting where required. 

For example, suppose an experiment reports the number concentrations in specified volume intervals $[a_{i},a_{i+1}]$ and the model being used solves the full discretized PBM, the corresponding model prediction $\hat{y}_{i}(t,\boldsymbol{\theta})$ for a given value of the model parameters $\boldsymbol{\theta}$ and time $t$ is
\begin{equation}
    \hat{y}_{i}(t,\boldsymbol{\theta}) = \int_{a_{i}}^{a_{i+1}} n(v,t,\boldsymbol{\theta}) dv.
\end{equation}
This mapping is declared using \texttt{DistributionObservable}:
\begin{minted}{julia}
measured_edges = [0.0, 1.0, 2.0, 4.0]

observable = DistributionObservable(
    measured_edges; coordinate=:volume,
)
\end{minted}
The \texttt{DistributionObservable} operator returns the predicted number concentration in the corresponding measurement interval. These intervals do not need to coincide with the discretized grid.  Moment-based predictions such as mean particle size / holdup can be defined using \texttt{MomentObservable}, and bulk predictions can be defined using \texttt{BulkObservable}.  The predictions obtained from the observation operators are then used to construct the likelihood. Assuming Gaussian measurement errors that are conditionally independent given $\boldsymbol{\theta}$, the log likelihood is
\begin{equation}
    \log{\mathcal{L}}(\boldsymbol{\theta}) = -\frac{1}{2}\sum_{i,j} \left( \frac{(y_{i,j} - \hat{y}_{i}(t_{j},\boldsymbol{\theta}))^{2}}{\sigma_{i,j}^{2}} + \log{\left(2\pi \sigma_{i,j}^{2} \right)}\right),
\end{equation}
where $y_{i,j}$ is the measurement data, $\sigma_{i,j}^{2}$ is the measurement variance, and the indices $i$ and $j$ index over the measurements and time respectively. This likelihood is implemented using \texttt{GaussianLikelihood}. A \texttt{Measurement} block combines the experimental measurements, their observation operator, and the selected likelihood:
\begin{minted}{julia}
measurement = Measurement(
    measurement_times, measured_values, observable;
    likelihood=GaussianLikelihood(sigma=measurement_std),
)
\end{minted}
Alternative likelihoods can be selected according to the type of data and assumed measurement errors. Several measurement blocks can be connected together and associated with one forward problem through \texttt{InferenceExperiment}, including population and bulk measurements from a coupled system. For the preceding example:
\begin{minted}{julia}
experiment = InferenceExperiment(
    :batch, problem, measurement;
    solver
)   
\end{minted}
Here, \texttt{:batch} is a user-selected identifier for the experimental run and the solver evaluates the forward problem at the required measurement times. Data from experiments conducted under different initial or operating conditions are represented by separate \texttt{InferenceExperiment} objects. These are combined into a single \texttt{EstimationProblem} to estimate shared parameters jointly across runs. The total log likelihood sums contributions across measurement blocks and experiments, assuming their measurement vectors are conditionally independent given the fitted parameters.   

Prior knowledge about the system is captured in the parameter estimation procedure through the inclusion of prior distributions for the unknown parameters. The package supports continuous univariate priors from \texttt{Distributions.jl} such as lognormal and uniform distributions. Each prior describes the physical parameter supplied to the PBM and is declared through \texttt{ParameterSpec}, together with a physical starting guess and an explicit transformation. For example, a growth rate parameter $g_{0}$ with a normal prior restricted to positive values and a logarithmic transformation for estimation purposes can be specified as:
\begin{minted}{julia}
using Distributions: Normal, truncated

growth_parameter = ParameterSpec(
    :g0;
    initial=0.2,
    prior=truncated(Normal(0.2, 0.05), 0.0, Inf),
    transform=LogTransform(),
)
\end{minted}
The prior is specified for $g_{0}$, while \texttt{LogTransform()} selects $z = \log g_{0}$ for optimization and posterior sampling to improve numerical scaling. For multiple fitted parameters, the specified transformations define the variables $\mathbf{z} = T(\boldsymbol{\theta})$ used for optimization and posterior sampling. The transformed prior density can therefore be written as
\begin{equation}
    \pi_{z}(\mathbf{z}) = \pi_{\theta}(T^{-1}(\mathbf{z}))\bigg| \det{\left(\frac{\partial T^{-1}(\mathbf{z})}{\partial \mathbf{z}}\right)} \bigg|.
\end{equation}
The package automatically accounts for the transformation in variables, so users need only specify priors in the physical parameter space. For the remainder of this section, the likelihood and posterior are expressed in terms of $\mathbf{z}$, with the physical parameters supplied to the forward model recovered through $\boldsymbol{\theta} = T^{-1}(\mathbf{z})$. MAP estimation minimizes the negative log posterior and can be written as
\begin{equation}
    \mathbf{z}^{*}_{\mathrm{MAP}} = \arg\min_{\mathbf{z}} \{-\log \mathcal{L}(\mathbf{z}) - \log{\pi_{z}}(\mathbf{z})\}.
\end{equation}
The parameter declarations and experiments are combined in an \texttt{EstimationProblem}, allowing parameters to be fitted jointly across experiments while retaining much of the machinery associated with the forward problem. MAP estimation is performed using \texttt{fit} through \texttt{Optim.jl} \cite{KMogensen2018}:
\begin{minted}{julia}
using Optim

estimation = EstimationProblem(
    (growth_parameter,), (experiment,),
)
map_result = fit(estimation; nstarts=5, seed=2022)
\end{minted}

The default optimizer used is BFGS with a backtracking line search. Gradients can be evaluated using \texttt{ForwardDiff} where compatible while optional forward / adjoint ODE sensitivities through \texttt{SciMLSensitivity.jl}, and explicitly selected finite differences can also be used. Multistart can be configured through \texttt{nstarts} or user-supplied starting points through \texttt{starts}. Additional starting points are generated as seeded Gaussian perturbations about the initial value of $\mathbf{z}$ are used. Parameter scaling, iteration limits, and convergence tolerances are also configurable. 

Local posterior uncertainty estimates are available through two options: \texttt{gauss\_newton\_approximation} and \texttt{laplace\_approximation}. Both represent the posterior near the MAP estimate, $\mathbf{z}_{\textrm{MAP}}^{*}$, by a joint Gaussian distribution $\mathcal{N}(\mathbf{z}_{\mathrm{MAP}}^{*},\boldsymbol{\Sigma})$. For Gaussian likelihoods with fixed measurement uncertainties, \texttt{gauss\_newton\_approximation} estimates the posterior covariance using the prediction sensitivities $S_{k,l} = \frac{\partial \hat{y}_{k}(\mathbf{z})}{\partial z_{l}} \bigg|_{\mathbf{z}_{\textrm{MAP}}^{*}}$, and the prior curvature \cite{Gunawan2003,Beck1977ParameterScience},
\begin{equation}
    \Sigma_{GN} = \left(\mathbf{S}^{\top} \mathbf{V}_{\epsilon}^{-1}\mathbf{S} - \nabla_{\mathbf{z}}^{2}\log \pi_{z}(\mathbf{z}_{\mathrm{MAP}}^{*}) \right)^{-1},
\end{equation}
where $k$ indexes predictions stacked across measured outputs, times, and experiments, and $l$ indexes the components of $\mathbf{z}$, $\mathbf{V}_{\epsilon}$ is the measurement-error covariance matrix which is diagonal with entries given by $\sigma_{i,j}^{2}$ assuming that the errors are conditionally independent given $\mathbf{z}$. The prior-curvature term reduces to the inverse of the prior covariance when the prior is Gaussian in $\mathbf{z}$. Alternatively, \texttt{laplace\_approximation} evaluates the full Hessian of the log posterior \cite{Bishop2006},
\begin{equation}
    \Sigma_{L} = \left( -\nabla^{2}_{\mathbf{z}} (\log \mathcal{L}(\mathbf{z}) + \log{\pi_{z}}(\mathbf{z}))\big|_{\mathbf{z} = \mathbf{z}_{\mathrm{MAP}}^{*}} \right)^{-1}.
\end{equation}
For Gaussian likelihoods, the Laplace approximation retains residual-weighted second derivatives of the predictions omitted by the Gauss--Newton approach. Both routines check MAP convergence, derivative consistency, and curvature before accepting the approximation. Local sensitivity / curvature diagnostics, likelihood / posterior profiles, and prior comparisons are available through \texttt{identifiability\_report}, \texttt{estimation\_profile}, and \texttt{prior\_sensitivity} respectively. These tools are intended to help users with identifiability analysis (see \cite{Wieland2021,Berliner2021} for additional information on parameter identifiability). 

Posterior sampling using MCMC is available through \texttt{sample\_posterior}, which uses the No-U-Turn Sampler (NUTS) through \texttt{AdvancedHMC.jl} \cite{Xu2020advancedhmc}. Sampling targets the same posterior in $\mathbf{z}$ as MAP estimation, providing joint draws for estimating marginal distributions, credible intervals, and parameter correlations. Continuing the preceding example: 
\begin{minted}{julia}
using AdvancedHMC, MCMCDiagnosticTools

posterior = sample_posterior(
    estimation; initial=map_result, seed=2022,
)
report = diagnostics(posterior)
\end{minted}
For convenience, the package provides configurable sampling defaults covering warmup, sampler adaptation, and diagnostic stopping, informed by practical MCMC guidance \cite{Vehtari2021,Hogg2018,Betancourt2017}. By default, four chains each undergo 500 warmup iterations, during which adaptation is performed, followed by batches of 250 retained draws. After at least 500 retained draws per chain, successful stopping requires two consecutive checks satisfying rank-normalized split $\hat{R} \leq 1.01$, bulk and tail effective sample sizes of at least 400 for every fitted parameter evaluated jointly across chains, and an energy Bayesian Fraction of Missing Information (BFMI) $\geq 0.3$ in each chain. Both checks also require finite Monte Carlo standard errors of the means and no retained divergences or maximum-tree-depth hits. Sampling terminates unsuccessfully if diagnostic checks detects divergence or maximum-tree-depth hits or if the sampling budget is exhausted before the checks pass all criteria. 

\subsection{Uncertainty Quantification}

Uncertainty quantification evaluates how uncertain parameters, initial conditions, and operating inputs affect model predictions. \texttt{UncertainInputs} specifies input distributions or supplied joint realizations, including posterior parameter draws from the parameter estimation workflow. \texttt{UQProblem} combines these inputs with an existing forward problem, a numerical solver, and observable outputs. The package currently supports Monte Carlo propagation and non-intrusive polynomial chaos expansion (PCE), both of which reuse the existing forward-simulation interfaces. 

Monte Carlo uncertainty propagation uses random realizations of uncertain inputs to estimate the statistics of model outputs \cite{Smith2024}. The distributions for the uncertain inputs are declared through \texttt{UncertainInputs} and are combined with the forward problem, solver, and requested observables in \texttt{UQProblem}. This problem is passed to \texttt{propagate} alongside \texttt{MonteCarlo} which specifies the ensemble size. The \texttt{propagate} call manages the complete ensemble calculation. The resulting ensemble is summarized through \texttt{uq\_summary} to estimate means, variances, pointwise quantiles, and user-defined event probabilities. For the preceding model, uncertainty in the growth parameter $g_{0}$ can be propagated to the first volume moment:
\begin{minted}{julia}
using Distributions: Uniform

inputs = UncertainInputs((g0=Uniform(0.15, 0.25),))
uq = UQProblem(
    problem, inputs;
    outputs=(m1=MomentObservable(1),),
    times=[0.0, t_final], method=SSPRK33(), dt=0.01,
)

n_mc = 500
mc = propagate(
    uq, MonteCarlo(n_mc; seed=2022); max_solves=n_mc,
)
mc_statistics = uq_summary(mc)
\end{minted}

Non-intrusive PCE constructs a polynomial approximation of how model outputs depend on uncertain inputs \cite{Mishra2026,Kim2013,Xiu2002}. For a set of independent uncertain parameters $\Theta$ with distributions specified through \texttt{UncertainInputs}, a model prediction, $\hat{y}(t,\Theta)$, such as concentration or a population moment is approximated as
\begin{equation}
    \hat{y}(t,\Theta) \approx \sum_{k=0}^{P}c_{k}(t)\psi_{k}(\Theta).
\end{equation}
The expansion contains $P+1$ polynomial basis functions up to the selected total degree. The coefficients $c_{k}(t)$ are deterministic and specific to each output and time. The basis functions are orthonormal under the input probability distribution with $\psi_{0}=1$. Consequently, the coefficients provide approximations to the output mean and variance \cite{Mishra2026} with $\mathbb{E}[\hat{y}(t,\Theta)] \approx c_{0}(t)$ and $\mathrm{Var}[\hat{y}(t,\Theta)] \approx \sum_{k=1}^{P}c_{k}^{2}(t)$. The package currently provides Hermite, Legendre, and Jacobi polynomials for inputs with normal, uniform, and beta distributions respectively with product bases for multiple independent inputs.  

The coefficients can be constructed using least-squares regression \cite{Luthen2021,Berveiller2006} by default or Gaussian-quadrature projection \cite{Kim2013}. For regression, \texttt{pce\_design} generates a user-specified number of training inputs using Latin hypercube sampling by default. Forward simulations at these inputs provide the data for fitting a separate coefficient set for each requested output and time \cite{Makrygiorgos2020,Petzke2020}. Polynomial basis evaluations form the regression matrix, while the corresponding predictions supply the target vectors. Before running the training simulations, the package checks and rejects regression designs that are rank deficient (e.g., insufficient training budget) or poorly conditioned. The least-squares problem is solved using SVD, and the factorization is reused across outputs and times. For the preceding example, a quadratic PCE in the single uncertain parameter $g_{0}$ is constructed using 16 training inputs: 
\begin{minted}{julia}
degree = 2
n_train = 16

design = pce_design(
    inputs; n=n_train, method=:lhs, seed=2023,
)

pce = fit_pce(
    uq, PolynomialChaos(degree);
    design, max_solves=n_train,
)
pce_statistics = uq_summary(pce)   
\end{minted}

\texttt{uq\_summary} returns the means, variances, and covariances of the PCE predictions. The same function can estimate uncertainty internals for each output and time and probabilities of specified events from predictions obtained using \texttt{evaluate\_pce} at additional input realizations, without further simulations of the full PBM. \texttt{validate\_pce} can be used to assess the PCE approximation errors against PBM simulations at inputs not included in the training and \texttt{refine\_pce} compares the specified polynomial degrees and training designs using these errors and the stability of the predicted means and variances.

\subsection{Extensibility and Verification}

The separation of the declaration of the physical terms in a PBM from the numerical methods used for their solution allows the package to accommodate new functional forms for the various terms and numerical schemes. New kinetic expressions and kernels can be supplied to existing term constructors readily. A new per-term discretization defines a method type and implements \texttt{prepare\_term} for applicability checks and precompute any required coefficients, and \texttt{add\_term!} to evaluate its contribution to the discretized PBM. To add a new moment method, users define a method type and implement its closure / quadrature reconstruction, together with any additional routines needed to evolve its variables and calculate population moments. The developer documentation specifies the required interfaces and provides extension examples. 

Automated unit tests check individual model terms and numerical methods. Numerical verification includes comparison with known analytical solutions and independent references calculations, conservation checks, and convergence tests under grid and timestep refinement.Integration tests then check that these components work together correctly in forward simulations, parameter estimation, and uncertainty quantification. For the latter two workflows, problems with known posterior distributions or output statistics provide reference results. Continuous integration reruns the suite to detect issues when existing implementations are modified or new functionality is incorporated. 

\clearpage

\section{Case Studies}

\subsection{Evaluating Numerical Schemes for Constant Growth}

Consider a 1-D PBM with constant growth rate and no source / sink terms with a Gaussian initial profile,
\begin{equation}
    \frac{\partial n}{\partial t} + 0.1\frac{\partial n}{\partial a} = 0, \quad n_{0}(a) = 100 \exp{\left(-\frac{(a-0.2)^{2}}{2(0.03)^{2}} \right)},
\end{equation}
where $n$ is the population density and $n_{0}$ is the initial condition.This model is equivalent to a constant velocity advection problem and has the analytical solution $n(a,t) = n_{0}(a-0.1t)$. The simplicity of this problem lends itself to benchmarking and validating solver implementations \cite{Inguva2022,Gunawan2004}. A domain $a\in[0,1]$ is considered with a final simulation time of 4s. Zero inflow flux and free outflow boundary conditions were applied at $a=0$ and $a=1$ respectively. Various growth term discretization and time integrators are considered with varying $\mathrm{CFL}$ numbers and grid resolution. 

\begin{figure}[ht]
    \centering
    \begin{subfigure}{0.8\textwidth}
        \centering
        \includegraphics[width=\textwidth]{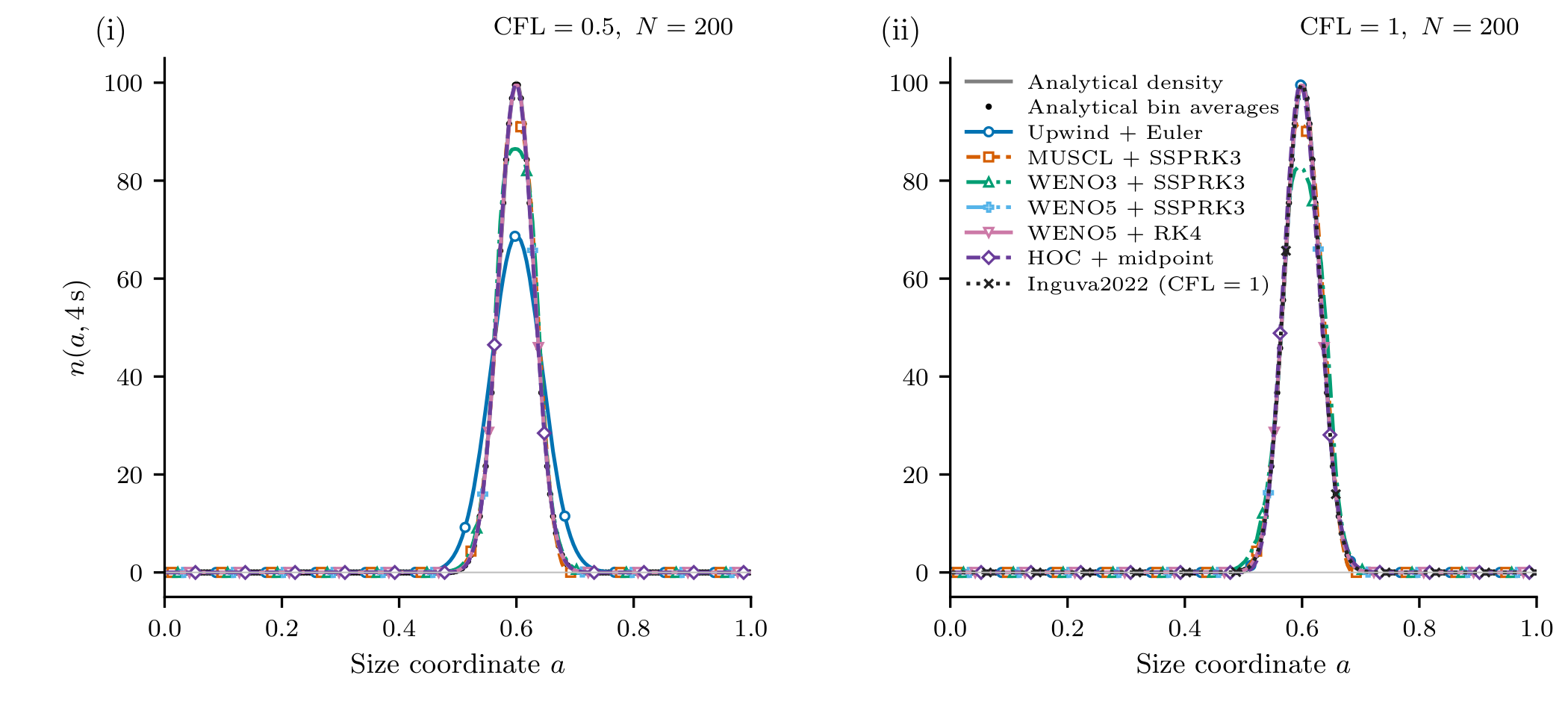} 
        \vspace{-0.6cm}
        \caption{Population distribution profiles.}
        \label{fig:case1a}
    \end{subfigure}
    \hfill
    \centering
    \begin{subfigure}{0.8\textwidth}
        \centering
        \includegraphics[width=\textwidth]{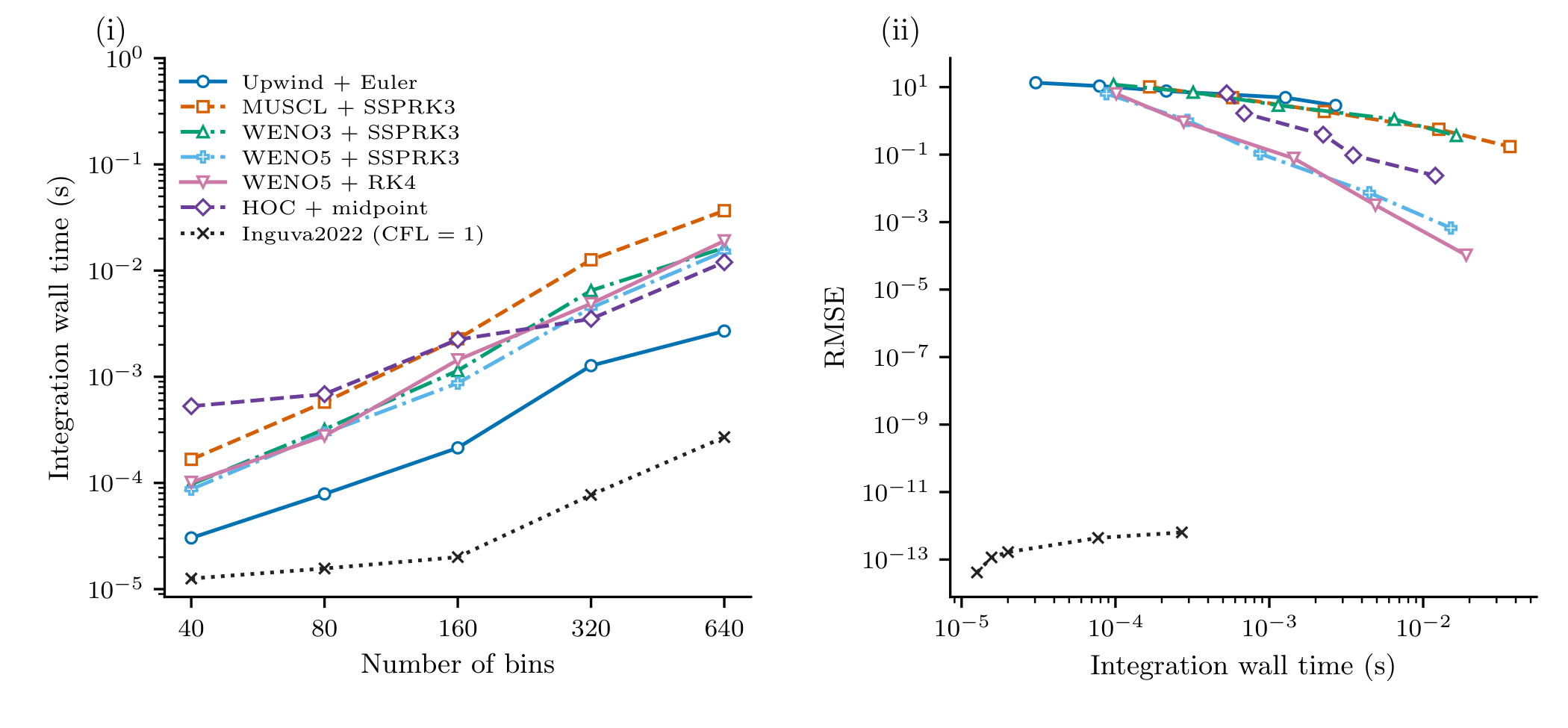} 
        \vspace{-0.6cm}
        \caption{Computational cost and accuracy}
        \label{fig:case1b}
    \end{subfigure}
    \caption{Comparison of different numerical methods for solving a constant-growth rate PBM with no source / sink terms. (a) Population distribution profiles using 200 uniformly spaced bins at (i) $\mathrm{CFL}=0.5$ and (ii) $1$. (b) Computational cost and accuracy at different grid resolutions: (i) integration wall time versus number of bins and (ii) RMSE versus integration wall time. All schemes use $\mathrm{CFL} = 0.5$ in (b) except \texttt{Inguva2022} which specifies $\mathrm{CFL} = 1$.}
    \label{fig:case1}
\end{figure}

The numerical solution for the various schemes using 200 uniform bins at $t=4\text{s}$ using $\mathrm{CFL} = 0.5$ and $1$ is shown in Fig.~\ref{fig:case1a} along with the reference analytical solution. At $\mathrm{CFL=0.5}$, the upwind scheme and forward Euler timestepping unsurprisingly performs the worst with significant numerical diffusion while other schemes perform better with WENO5 and HOC performing the best. However, at $\mathrm{CFL}=1$, upwind with forward Euler solves the PBM to machine precision as each timestep translates the profile by exactly one bin \cite{Inguva2022}. Computational cost and accuracy are also compared using on grids of varying resolution (Fig.~\ref{fig:case1b}), with $\mathrm{CFL}=0.5$ for all schemes except \texttt{Inguva2022} which requires $\mathrm{CFL}=1$. Upwind with forward Euler is comparatively computationally cheap, but retains the largest error under grid refinement. Among the higher-order methods considered, WENO5 performs the best, achieving the smallest errors at comparable computational costs. The \texttt{Inguva2022} scheme is an outlier in that it is able to solve the problem to machine precision at significantly lower costs, illustrating the benefit of selecting a numerical scheme that exploits the mathematical structure of the model when possible. It should also be noted that the comparisons presented here reflect the specific model and solver configurations used. For example, the HOC-ADI scheme permits using significantly larger time-steps ($\mathrm{CFL}>1$) \cite{Li2026}, which could provide improved error--efficiency performance.

\subsection{Isothermal Batch Protein Crystallization}

This case study considers the crystallization of lysozyme from an initially unseeded, well-mixed 100mL batch system at 20$^{\circ}$C with an initial protein concentration of $c_{0} = 20 \text{g/L}$. Experimental data was obtained from \cite{Tang2018} and comprised the dissolve protein concentration profile and the number distributions of equivalent crystal diameter at 74, 180, 420, and 660min. This system was previously modeled by \cite{Mitchell2023} who used the dissolved protein concentration profile and the median crystal sizes for parameter estimation. 

The present case study instead uses the crystal-size distributions over the diameter range $2.5$-$25 \,\mu\mathrm{m}$ for parameter estimation. The model couples the PBM for the crystal number density $n(L,t)$ (where the equivalent diameter, $L$, is the intrinsic coordinate) to the bulk dissolved protein concentration, $c(t)$. The PBM incorporates primary nucleation (using classical nucleation theory (CNT)), and size-independent growth (using a power-law expression). Additionally, growth rate dispersion (GRD) is incorporated to account for further distribution broadening through variability in individual crystal growth rates, a phenomena previously observed in lysozyme \cite{Judge2010}. The size-coordinate dispersion coefficient is represented phenomenologically as a function of the growth rate i.e., $D = k_{D} G$ \cite{Makoveeva2018}, where $k_{D}$ is a model parameter. Aggregation, breakage, and dissolution are neglected. Nucleation introduces a total number flux $J$ at the lower boundary using a negligible birth-size approximation, and zero flux is imposed at the upper boundary $L_{\max} = 50 \mu\text{m}$. The model is given by,
\begin{equation}
\begin{gathered}
\frac{\partial n}{\partial t} + \frac{\partial (Gn)}{\partial L} =
\frac{\partial^{2}\left(Dn\right)}{\partial L^{2}},
\quad 0<L<L_{\max}, \\[2mm]
\frac{dc}{dt} = -k_v\rho\int_{0}^{L_{\max}} L^{3}\frac{\partial n}{\partial t}\,dL, \\[2mm]
G = A_G(S-1)^g, \quad J = A_J\exp\left(-\frac{16\pi\gamma^{3}\nu^{2}}{3(k_BT)^{3}(\ln S)^{2}}\right), \quad D = k_DG, \\[2mm]
\bigg(
Gn-D\frac{\partial n}{\partial L} \bigg|_{L=0} =J,
\quad
\bigg(
Gn-D\frac{\partial n}{\partial L} \bigg|_{L=L_{\max}} =0, \\[2mm]
n(L,0)=0, \quad c(0)=c_0.
\end{gathered}
\label{eq:protein_pbm}
\end{equation}
Here, $G$ and $J$ are the crystal growth and nucleation rates respectively and $S$ is the supersaturation ratio given by $S = \frac{c}{c^{*}}$. Where $S \leq 1$, the nucleation, growth, and dispersion terms are set to zero. The parameters $A_{G}$ and $A_{J}$ are the corresponding kinetic prefactors, $g$ is the growth exponent, and $k_{D}$ is the GRD constant. The crystal density $\rho$ and shape factor $k_{v}$ relate the crystal size to crystal mass. The CNT expression involves the crystal-solution interfacial energy $\gamma$, the volume per protein molecule $v$, the Boltzmann constant $k_{B}$, and the temperature $T$. Values for the various model constants were obtained from \cite{Mitchell2023}. Four parameters, $A_{J}$, $A_{G}$, $g$, and $k_{D}$ were estimated jointly with priors based on the parameter ranges reported in \cite{Mitchell2023}. Specifically, $\log{A_{J}} \sim \mathcal{U}(-2.8134, 97.1866)$, and $\log{A_{G}} \sim \mathcal{U}(\log{10^{-5}}, \log{1.5})$ with units of $\text{L}^{-1}\text{min}^{-1}$ and $\mu\text{m} \ \text{min}^{-1}$ for $A_{J}$ and $A_{G}$ respectively. An expanded range for the growth exponent of $g\sim(1,5)$ was used and an exponential prior with a mean of 0.1$\mu \text{m}$ was used for $k_{D}$ based on a scaling estimate of the data. The interfacial energy $\gamma$ was fixed at 0.64 $\text{mJ} \ \text{m}^{-2}$. 

\begin{figure}[ht]
    \centering
    \begin{subfigure}{0.8\textwidth}
        \centering
        \includegraphics[width=\textwidth]{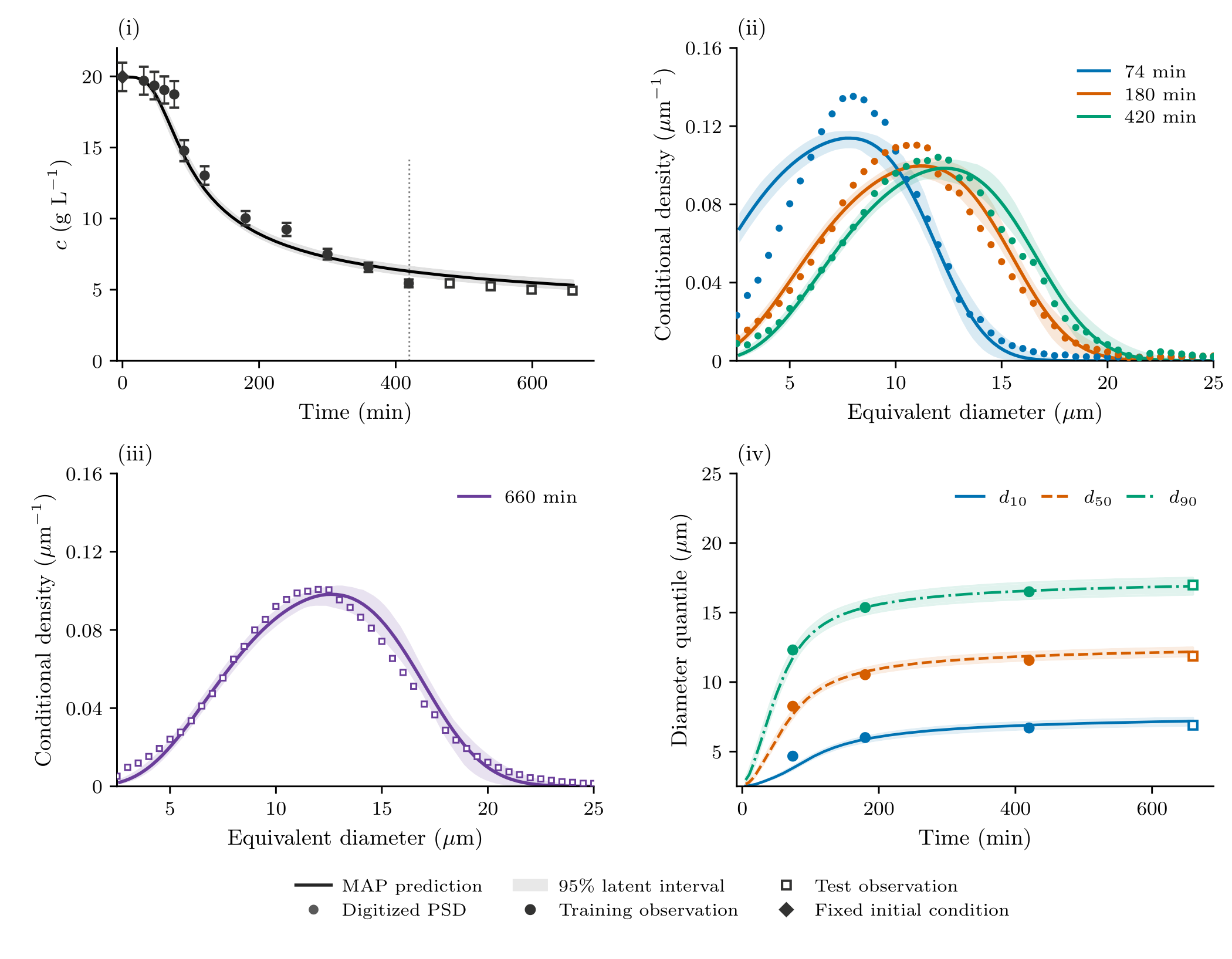} 
        \vspace{-0.6cm}
        \caption{Model fit and test predictions.}
        \label{fig:case2a}
    \end{subfigure}
    \hfill
    \centering
    \begin{subfigure}{0.8\textwidth}
        \centering
        \includegraphics[width=\textwidth]{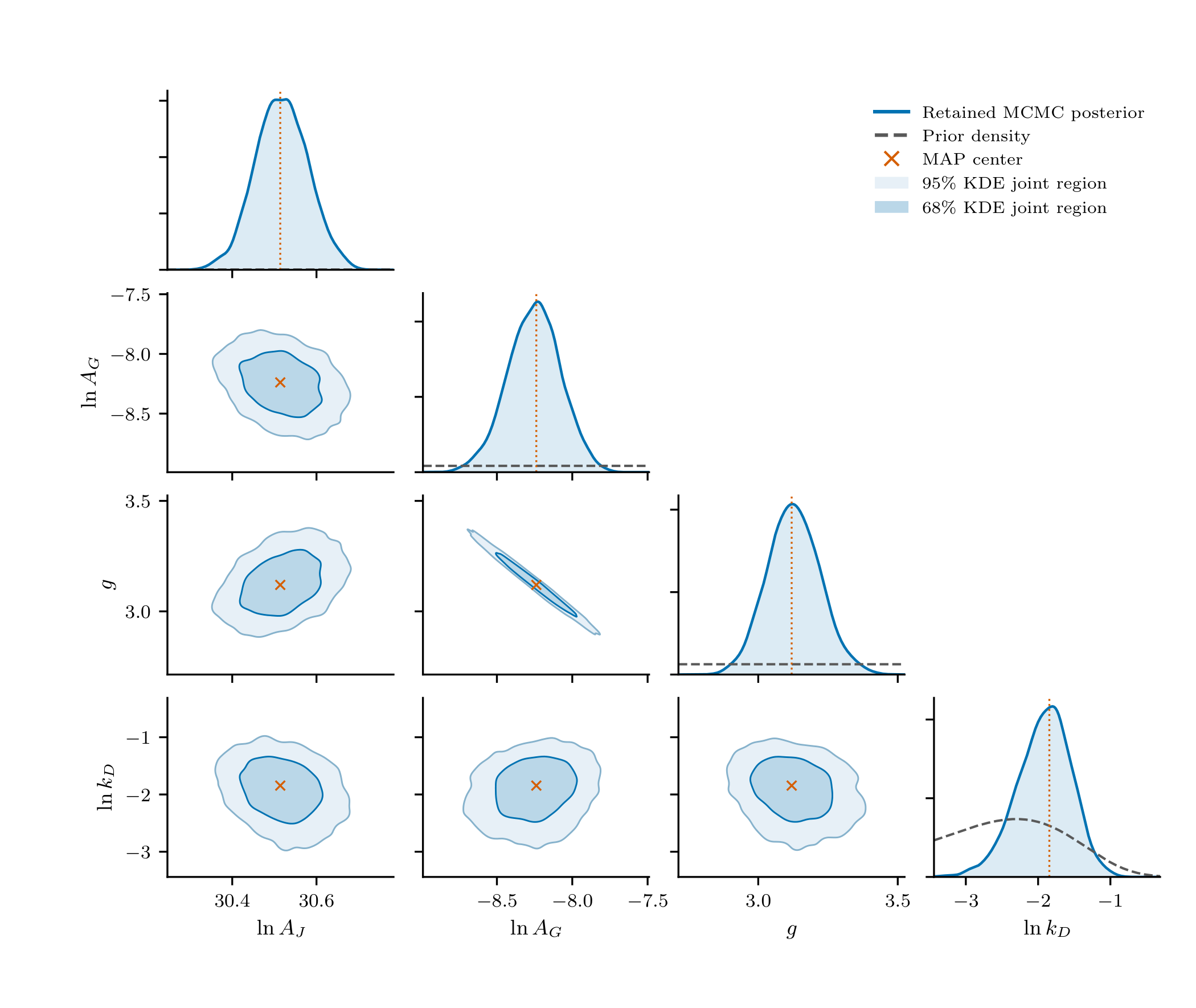} 
        \vspace{-0.6cm}
        \caption{Marginal and joint posterior parameter distributions.}
        \label{fig:case2b}
    \end{subfigure}
    \caption{Bayesian parameter estimation for the isothermal batch lysozyme crystallization case study with data obtained from \cite{Tang2018}. The model incorporated CNT nucleation, power-law growth, and growth-rate dispersion. (a) (i) MAP predictions for  bulk protein concentration, (ii) particle-size distributions used for parameter estimation, (iii) test particle-size distribution at 660 min, and (iv) evolution of particle size percentiles $d_{10}, d_{50},$ and $d_{90}$. (b) Marginal MCMC posterior and prior densities (on the diagonal plots) with the 68\% and 95\% regions of the pairwise posterior regions. Orange lines and crosses mark the MAP estimate.}
    \label{fig:case2}
\end{figure}

The model was calibrated using concentration and size distribution data up to 420 min, with the later concentration measurements and the size distribution at 660 min used for testing. The MAP predictions were able to capture the overall concentration profile, the evolution of 10th, 50th, and 90th percentiles of the particle-size distribution ($d_{10},d_{50}$, and $d_{90}$), and the particle-size distributions at later times reasonably well (see Fig.~\ref{fig:case2a}). Some discrepancy remains in the early-time (74 min) particle-size distribution, especially with the peak height. For comparison, the same model without the GRD term was also tested (this can be found in the package documentation). Although the model was able to reproduce the evolution of the bulk protein concentration and particle-size distribution percentiles similarly well, substantial discrepancies in the predicted distribution shapes persisted for all times suggesting a limitation of this reduced model structure in describing the process. The posterior distributions are substantially narrower than the priors indicating that the data reasonably constrains the parameters, though a negative correlation remains between $A_{G}$ and $g$ which reflects the compensation between the growth prefactor and exponent (see Fig.~\ref{fig:case2b}). Overall, this case study demonstrates the implementation of a PBM coupled to a bulk balance and the parameter estimation functionalities of the package. 

\subsection{Axially-Dispersed Plug Flow Precipitator}

This case study considers the isothermal continuous precipitation in an axially-dispersed plug flow reactor, following Case 7a of \cite{Zhang2024}. The system assumes a feed initially free of any particles and a dissolved solute concentration of 9$\text{kg} \ \text{m}^{-3}$ that enters a 0.47m tubular reactor. The PBM incorporates primary and secondary nucleation with size-independent growth, coupled to a bulk solute balance. Both the particles and bulk solute undergo advection and dispersion, while aggregation, breakage, and other effects are neglected. The model equations are
\begin{equation}
\begin{gathered}
\frac{\partial n}{\partial t}
+ v_{\mathrm{ax}}\frac{\partial n}{\partial z}
+ \frac{\partial(Gn)}{\partial x}
= D_{\mathrm{ax}}\frac{\partial^{2}n}{\partial z^{2}},
\\[2mm]
\frac{\partial c}{\partial t}
+ v_{\mathrm{ax}}\frac{\partial c}{\partial z}
= D_{\mathrm{ax}}\frac{\partial^{2}c}{\partial z^{2}}
- \rho k_v\left(
B_0x_c^{3}
+ 3G\int_{x_c}^{x_{\max}}x^{2}n\,dx
\right),
\\[2mm]
G = k_gs, \quad
B_0 = k_ps^{u} + k_bs^{b}M,
\\[2mm]
\left.Gn\right|_{x=x_c} = B_0, \quad
\left.Gn\right|_{x=x_{\max}} = 0,
\\[2mm]
\bigg( v_{\mathrm{ax}}n
- D_{\mathrm{ax}}\frac{\partial n}{\partial z}\bigg|_{z=0} = 0, \quad
\bigg(
v_{\mathrm{ax}}c
- D_{\mathrm{ax}}\frac{\partial c}{\partial z}
\bigg|_{z=0}
= v_{\mathrm{ax}}c_{\mathrm{feed}},
\\[2mm]
\left.\frac{\partial n}{\partial z}\right|_{z=L} = 0,
\quad
\left.\frac{\partial c}{\partial z}\right|_{z=L} = 0,
\\[2mm]
n(x,z,0) = 0, \quad c(z,0) = 0.
\end{gathered}
\label{eq:precipitation_pbm}
\end{equation}
Here $n(x,z,t)$ is the particle number density, $c(z,t)$ is the bulk concentration, and $s = \max{(\frac{c}{c^{*}}-1, 0)}$ is the relative supersaturation. Secondary nucleation depends on the local crystal mass concentration $M = \rho k_{v}\int_{x_{c}}^{x_{\max}} x^{3} n dx$, where $\rho$ and $k_{v}$ are the particle density and volume shape factor. The bulk solute balance accounts for material consumed by both nucleation and growth. Danckwerts and zero dispersive flux boundary conditions were applied to the inlet and outlet respectively. The model and simulation parameters can be found in \cite{Zhang2024}.

\begin{figure}[ht]
    \centering
    \begin{subfigure}{0.8\textwidth}
        \centering
        \includegraphics[width=\textwidth]{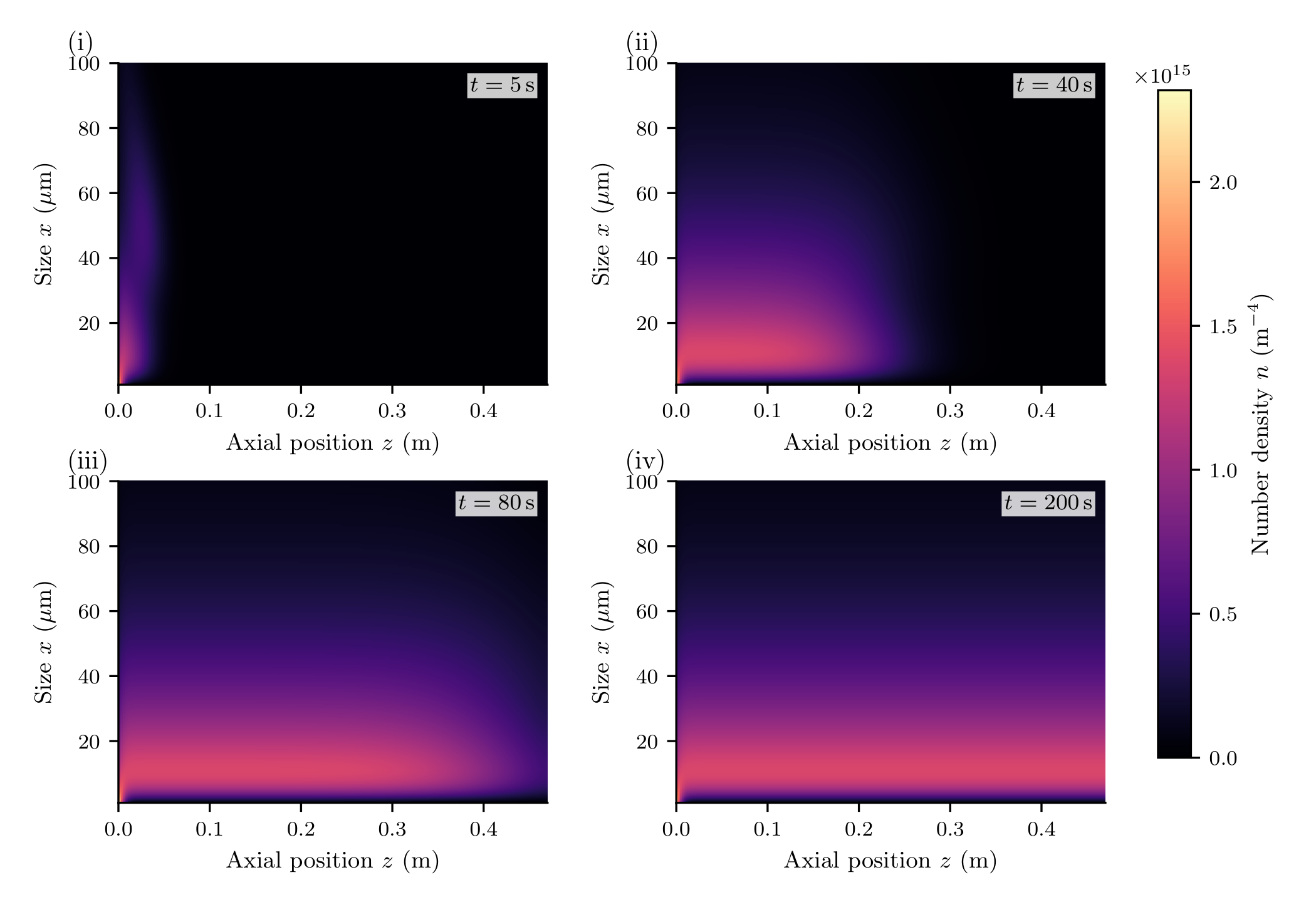} 
        \vspace{-0.6cm}
        \caption{Transient population density fields.}
        \label{fig:case3a}
    \end{subfigure}
    \hfill
    \centering
    \begin{subfigure}{0.8\textwidth}
        \centering
        \includegraphics[width=\textwidth]{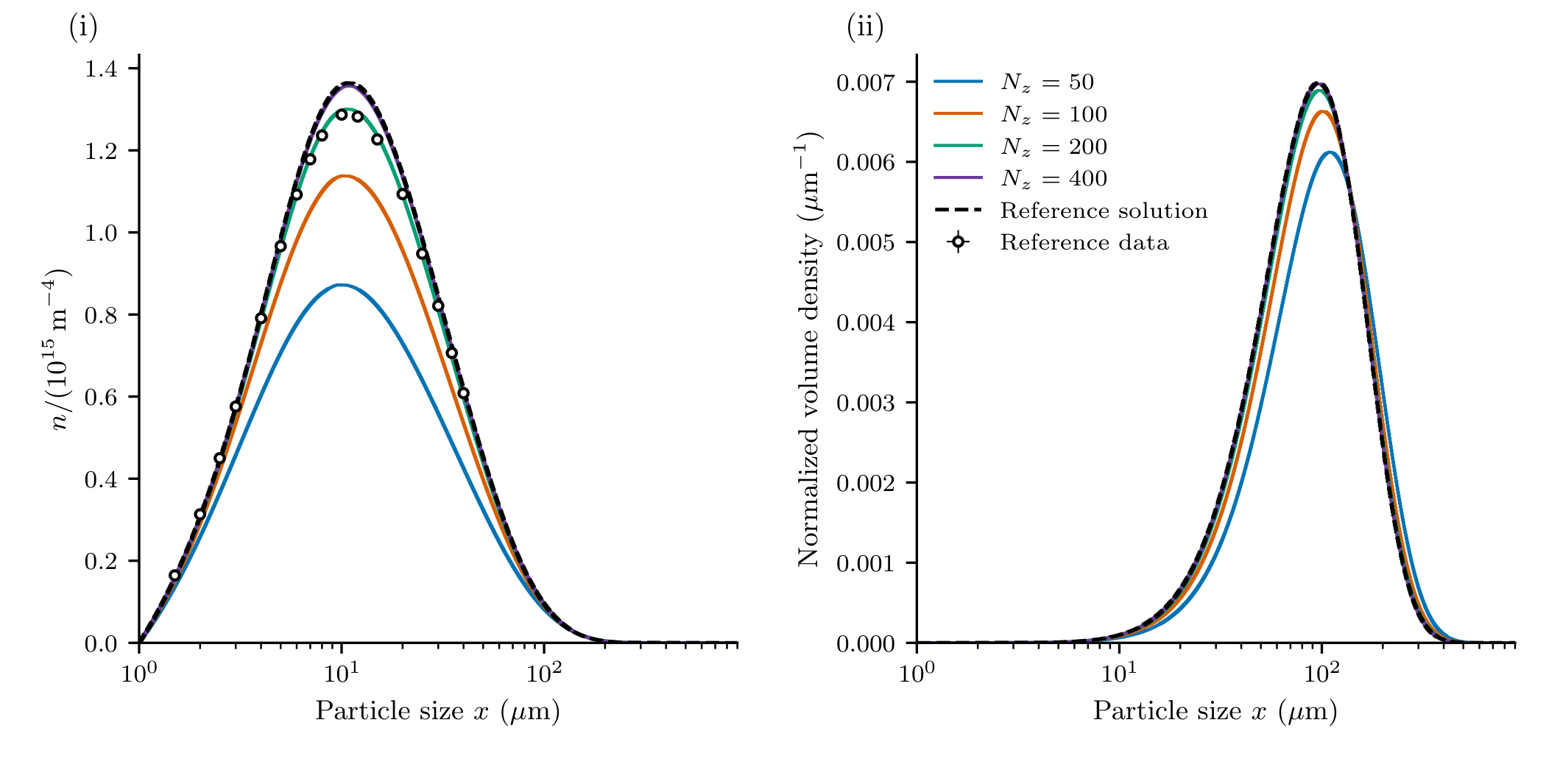} 
        \vspace{-0.6cm}
        \caption{Outlet distributions under mesh refinement.}
        \label{fig:case3b}
    \end{subfigure}
    \caption{Continuous precipitation in an axially-dispersed plug flow reactor based on the benchmark case 7a in \cite{Zhang2024}. (a) Population density fields at 5, 40, 80, and 200s using 1000 size bins and 500 axial cells. (b) (i) Outlet number density and (ii) normalized volume density at $t=200\text{s}$ for different axial grid resolutions. The dashed curves denote the reference numerical solution generated in this work and circles correspond to reference data of the same benchmark obtained from \cite{Zhang2024}.}
    \label{fig:case3}
\end{figure}

The PBM and bulk solute balances are coupled using \texttt{PBMSystem}, with \texttt{axial\_problem} adding the reactor transport. Growth is discretized using the \texttt{WENO5} scheme on logarithmically spaced bins and spatial advection is discretized using the \texttt{MUSCL} advection scheme on a uniformly spaced grid. Time integration is carried out using the external CVODE-BDF solver. A reference numerical solution (see Fig.~\ref{fig:case3}) was generated using 1000 size bins and 500 axial cells and the axial refinement simulations (see Fig.~\ref{fig:case3b}) retained the same size grid and used 50-400 axial cells. Reference data was also obtained from \cite{Zhang2024} on the outlet number density profile, though this was at a lower size and axial resolution. The population density fields (see Fig.~\ref{fig:case3a}) showcase expected physical behavior with particles forming near the inlet and progressively occupying the reactor as the feed advances. The outlet distributions at $t=200\text{s}$ demonstrate the importance of grid resolution with coarser axial meshes under-predicting the distribution peaks. The number distributions also agree reasonably with the profile obtained from \cite{Zhang2024}, indicating that the model implementation in this package was able to reproduce the benchmark. 

\subsection{Uncertainty Quantification of a Milling Process}

This case study considers continuous conical screen milling using the model developed by \cite{Barrasso2013} and subsequently considered for uncertainty quantification by \cite{Bhonsale2018}. Particle volume $v$ is used as the intrinsic coordinate and $n(v,t)$ represents the number of particles held up in the mill. The population balance model accounts for feeding, breakage, and discharge through the screen as follows,
\begin{equation}
\begin{gathered}
\frac{\partial n(v,t)}{\partial t}
= F_{\mathrm{in}}(v)-F_{\mathrm{out}}(v,t)
+ B(v,t)-S(v,t)n(v,t),
\\[2mm]
B(v,t)
= \int_v^\infty b(v,y)S(y,t)n(y,t)\,dy,
\\[2mm]
F_{\mathrm{out}}(v,t)
= \left(1-f_d[d(v)]\right)
\left[F_{\mathrm{in}}(v)+B(v,t)-S(v,t)n(v,t)\right],
\\[2mm]
n(v,0)=n_0(v), \quad v>0.
\end{gathered}
\label{eq:milling_pbm}
\end{equation}
Here, $F_{\mathrm{in}}$ and $F_{\mathrm{out}}$ are the number densities supplied by the feed and discharged through the screen per unit time respectively. $S$ is the breakage frequency of a particle with volume $v$, and $b$ is the daughter number density. The constitutive expressions for $F_{\mathrm{in}}$, $S$, and $b$, together with the screen retention function $f_{d}$ are given by,
\begin{equation}
\begin{gathered}
S(v,t)
= \alpha\Omega(t)
\left(\frac{v}{v_{\mathrm{ref}}}\right)^\gamma,
\\[2mm]
b(v,y)
= \begin{cases}
\displaystyle
\frac{C}{v\sigma}
\exp\left[
-\frac{\left(\ln(\kappa v/y)\right)^2}{2\sigma^2}
\right],
& 0<v\le y, \\[2mm]
0, & v>y,
\end{cases}
\\[1mm]
\int_0^y u\,b(u,y)\,\mathrm{d}u=y,
\qquad y>0,
\\[2mm]
f_d(d)
= \begin{cases}
0, & d\le(1-\delta)d_{\mathrm{screen}}, \\[1mm]
\displaystyle
\frac{d-(1-\delta)d_{\mathrm{screen}}}
{\delta d_{\mathrm{screen}}},
& (1-\delta)d_{\mathrm{screen}}<d\le d_{\mathrm{screen}},
\\[2mm]
1, & d>d_{\mathrm{screen}},
\end{cases}
\\[2mm]
d(v)=\left(\frac{6v}{\pi}\right)^{1/3},
\quad
F_{\mathrm{in}}(v)
= \frac{\dot m_{\mathrm{in}}}{\rho v}f_{\mathrm{in}}(v),
\quad
\int_0^\infty f_{\mathrm{in}}(v)\,\mathrm{d}v=1.
\end{gathered}
\label{eq:milling_constitutive}
\end{equation}
Here, $\Omega(t)$ is the impeller speed, $v_{\mathrm{ref}}$ is a reference particle volume, and $\alpha$ and $\gamma$ are model parameters. The truncated lognormal distribution of the daughter-to-parent volume ratio has a median $\frac{1}{\kappa}$ and a logarithmic standard deviation of $\sigma$. $C$ is a normalization constant that ensures the total expected volume of the fragments from one breakage event is equal to the parent volume $y$. At an equivalent diameter $d(v)$, the screen retains a fraction $f_{d}$ of the feed and net breakage production. The aperture of the screen is $d_{\mathrm{screen}}$ and $\delta$ controls the width of the transition from full discharge to full retention. The feed term uses the mass feed rate $\dot{m}_{\mathrm{in}}$, particle density $\rho$, and a normalized volume-weighted feed distribution $f_{\mathrm{in}}$. The model parameters and feed characteristics follow \cite{Bhonsale2018,Barrasso2013}. 

Two comparisons are considered. The first compares nominal predictions of mill hold-up and outlet median particle diameter $d_{50}$ with the simulations in \cite{Barrasso2013} at constant impeller speeds of 3350 and 4923 rpm. The initial inventory is set to one second of feed (7.4 g). The second compares Monte Carlo and first- and second-degree PCE estimates of uncertainty in $d_{50}$ following speed reduction as considered by \cite{Bhonsale2018}. The mill starts empty, and the impeller speed drops from 4923 to 1500 rpm at 30s. Four model parameters are treated as uncertain: $\delta$, $\alpha$, $\gamma$, and $\sigma$. Each is assigned a normal distribution with the mean and standard deviation following \cite{Bhonsale2018} and the four parameters are assumed to be mutually independent. Both comparisons use the same numerical configuration of 100 geometrically spaced volume pivots, integration of the log-normal feed over each bin, conservative fixed-pivot breakage, and CVODE-BDF time integration. The mass feed rate is distributed among the bins in proportion to the integral of the lognormal feed distribution over each bin. 

\begin{figure}[ht]
    \centering
    \begin{subfigure}{0.8\textwidth}
        \centering
        \includegraphics[width=\textwidth]{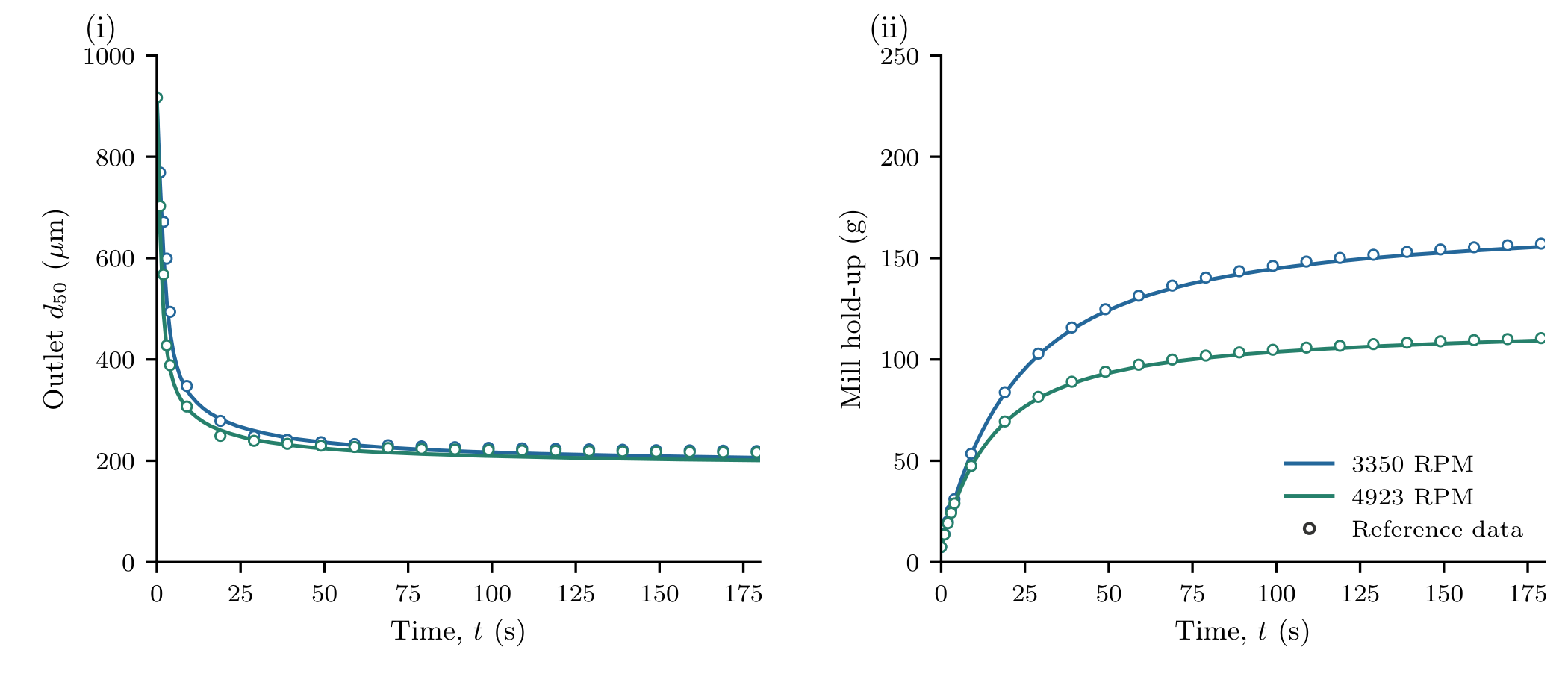} 
        \vspace{-0.6cm}
        \caption{Startup at constant impeller speed following \cite{Barrasso2013}.}
        \label{fig:case4a}
    \end{subfigure}
    \hfill
    \centering
    \begin{subfigure}{0.8\textwidth}
        \centering
        \includegraphics[width=\textwidth]{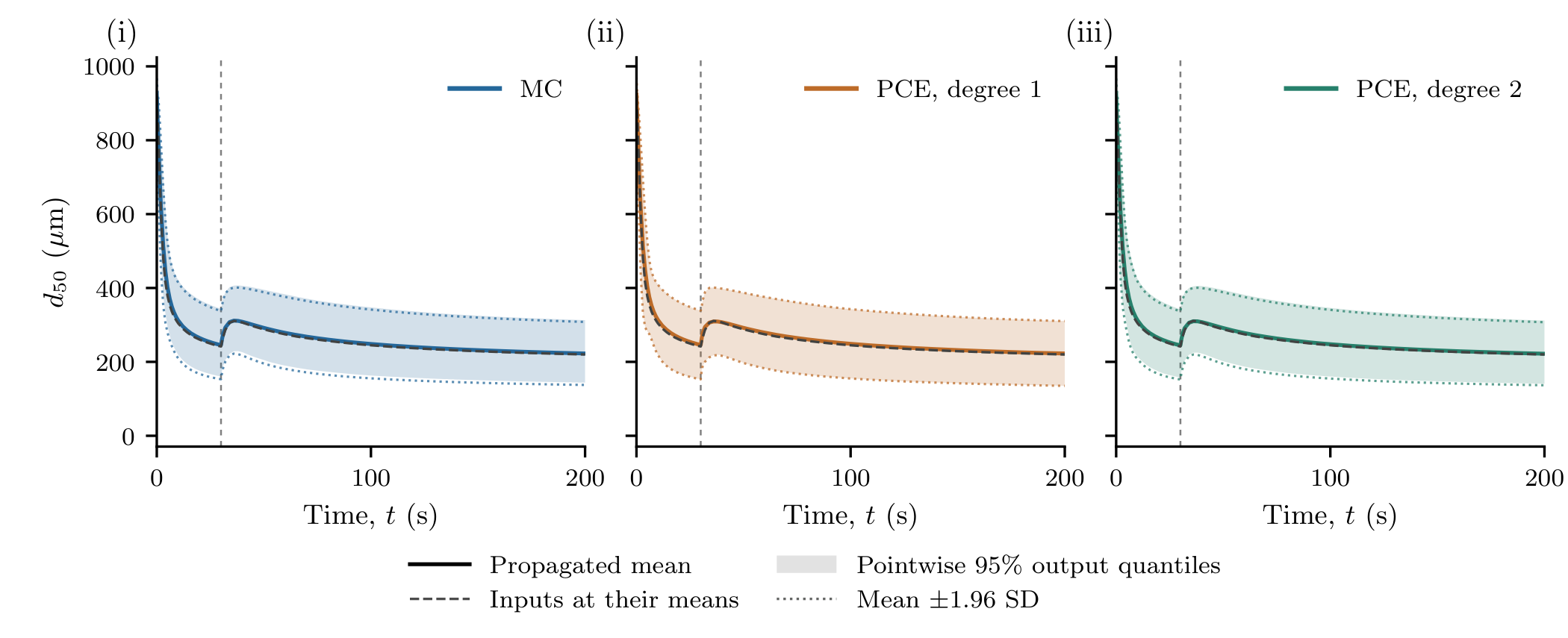} 
        \vspace{-0.6cm}
        \caption{Uncertainty in the outlet median particle size after speed reduction \cite{Bhonsale2018}.}
        \label{fig:case4b}
    \end{subfigure}
    \caption{Population balance modeling of a continuous conical screen milling process. (a) (i) Outlet $d_{50}$ and (ii) mill hold-up at 3350 and 4923 rpm. Markers show simulation results obtained from \cite{Barrasso2013}. (b) Uncertainty in the outlet $d_{50}$ following \cite{Bhonsale2018}, estimated using (i) Monte Carlo, (ii) first-degree PCE, and (iii) second-degree PCE. The solid curves show the propagated means and the shaded bands show the point-wise central 95\% output intervals. The dotted curves show the 95\% bounds of a normal approximation using the propagated mean and variance at each time and the dashed curves show the model predictions at the parameter means.}
    \label{fig:case4}
\end{figure}

The nominal predictions of mill hold-up agree closely with the simulations of \cite{Barrasso2013}, while the predicted outlet $d_{50}$ is slightly lower (see Fig.~\ref{fig:case4a}). Grid refinement reduces the error in $d_{50}$ and it has been previously noted that the simulation of this model is sensitive to the numerics \cite{Bhonsale2019}. The nominal comparison provides a check for the forward solver implementation and the current implementation is adequate for evaluating the uncertainty quantification workflow. A more detailed assessment of the numerics is beyond the scope of this case study. Using the same numerical configuration, both first- and second-degree PCE (which use 12 and 32 training simulations respectively) surrogates capture the transient increase in $d_{50}$ following speed reduction, and approximate the mean and pointwise 95\% intervals of the Monte Carlo reference (see Fig.~\ref{fig:case4b}). It should be noted that the predicted mean of $d_{50}$ is consistently higher than that reported in \cite{Bhonsale2018}. Differences in the numerics may contribute to this discrepancy, though the exact cause is unresolved. Despite this, the agreement between the PCE and Monte Carlo approaches demonstrates the uncertainty quantification workflow of the package.

\section{Conclusions and Roadmap}

\texttt{PopulationBalanceModels.jl} provides an open-source framework for implementing PBMs. It builds on Julia's performant numerical computing functionalities and capable scientific software ecosystem to support essential tasks  such as simulation, Bayesian parameter estimation, and uncertainty quantification. The package is designed such that model definition is kept separate from the numerical methods so that users can implement a PBM in an accessible manner that closely follows the structure of the model, while retaining full control of the numerics. The package currently supports a diverse range of model types, such as coupled PBMs, multi-dimensional PBMs, PBMs coupled to bulk mass/energy balances, and PBMs defined in an axially-dispersed plug flow reactor, and also a majority of the common mechanistic model terms used in PBMs. These functionalities are supported by a suite of natively implemented numerical methods that are commonly used for solving the various terms in a PBM using either direct discretization of the PDE or a moment representation. Extensibility in terms of either incorporating new mechanistic terms or numerical methods is supported with detailed guidance provided in the documentation. The current functionality of the package is demonstrated through the various case studies and examples presented in this work and also in the documentation. 

Future development of the package will take place along multiple axes and would be motivated by community uptake, feedback, and specific new applications that require new functionality. Some potential directions include: 
\begin{itemize}
    \item \textbf{New PBM formulations} Support for other classes of PBMs such as stochastic PBMs, mixed discrete--continuous PBMs, and PBMs coupled to algebraic constraints such as chemical equilibria would expand the scope of the package. 
    \item \textbf{Spatial transport and CFD coupling} Spatial transport is currently limited to axial advection and dispersion in tubular reactors. One-way coupling of the PBM in multidimensional spatial domains would require a spatial discretization for population transport, and an interface for importing fields from an external CFD solver such as velocity for advection, and composition for evaluating local kinetics. Two-way coupling of the PBM with solvers such as OpenFOAM could leverage frameworks such as preCICE \cite{Chourdakis2022} to manage the coupling. 
    \item \textbf{Hybrid modeling.} Hybrid PBMs retain the mechanistic balance equation structure, but use data-driven components to represent poorly characterized kinetic processes or approximate costly calculations such as aggregation \cite{Pessina2026,Shah2026}. Julia's scientific machine learning ecosystem provides the tools to implement this capability. 
    \item \textbf{Model-based process development and operation.} Incorporating workflows from process systems engineering would support the development of predictive PBMs and their subsequent use in process design and control \cite{Destro2024}. Model-based design of experiments (MBDoE) can help guide the selection of experiments that distinguish between competing models or improve the precision of parameter estimates. Dynamic optimization can use a validated PBM to determine optimal operating policies. These capabilities could reuse the current simulation and observation functionality, with optimization and control routines requiring additional development work for provision within the package either natively or through interfaces to other packages.
    \item \textbf{Advanced numerical methods and computational efficiency.} Further developments to the numerical methods supported by the package natively give users more options for solving a PBM. Depending on the problem, specific methods may be more performant. New schemes also help expand the range of supported models types such as aggregation / breakage for multidimensional PBMs which is currently not supported. Additionally, further optimization of the current code base and existing routines can help reduce the cost of individual simulations, with particular benefit for workflows that require repeated forward solves.
\end{itemize}

It is hoped that with the introduction of \texttt{PopulationBalanceModels.jl}, users considering PBMs for new applications will be able to build on a common platform with tested solvers and analysis tools. As AI agents are taking on a larger role in building and running scientific and engineering models, they could use the same documented interfaces as researchers, keeping modeling assumptions and numerical choices explicit for review. We welcome contributions of model implementations, numerical methods, and benchmark problems so that advances made for one application can be assessed and built upon in others.

\section*{Code Availability}

The repository and package installation instructions can be found at \url{https://github.com/pavaninguva/PopulationBalanceModels.jl}. The documentation and worked case studies can be found at \url{https://pavaninguva.github.io/PopulationBalanceModels.jl/dev/}

\section*{Acknowledgments}
This research did not receive any specific grant from funding agencies in the public, commercial, or not-for-profit sectors.

\section*{Declaration of Generative AI Use}
The development of the code base for the package utilized OpenAI Codex (GPT 6.1 Sol) as a programming assistant. ChatGPT (GPT 6.1 Sol) was used to digitize the various plots to obtain the reference data present in case studies 2-4. Outputs were checked manually after extraction.

\clearpage
\printbibliography

\end{document}